\documentclass[%
 reprint,
superscriptaddress,
nobibnotes,
 amsmath,amssymb,
 aps,
pre,
prstper,
floatfix,
]{revtex4-1}

\usepackage{graphicx}
\usepackage{dcolumn}
\usepackage{bm}
\usepackage{hyperref}
\usepackage[mathlines]{lineno}
\usepackage{xcolor}

\usepackage{appendix}

\begin{document}

\preprint{APS/123-QED}

\title{ \textcolor{black}{Variational approach} to study soliton dynamics in a passive fiber loop resonator with coherently driven phase-modulated external field}

\author{Maitrayee Saha}
 \email{maitrayee@iitkgp.ac.in}
 \affiliation{
 Department of Physics, Indian Institute of Technology Kharagpur, Kharagpur-721302, India
 }
 \author{Samudra Roy}
 
 \affiliation{
 Department of Physics, Indian Institute of Technology Kharagpur, Kharagpur-721302, India
 }
 \author{ Shailendra K. Varshney}

 \renewcommand{\andname}{\ignorespaces}
 
\affiliation{
 Department of Electronics and Electrical Communication Engineering, Indian Institute of Technology Kharagpur, Kharagpur-721302, India
}

\date{\today}

\begin{abstract}
We report a detailed semi-analytical treatment to investigate the dynamics of a single cavity soliton (CS) and two co-propagating CSs separately in a Kerr mediated passive optical fiber resonator which is driven by a phase-modulated pump. The perturbation is dealt with by introducing a Rayleigh$^,$s dissipation function in the framework of variational principle that results in a set of coupled ordinary differential equations describing the evolution of individual soliton parameters. We further derive closed form expressions for quick estimation of the temporal trajectory, drift velocity and the phase shift accumulated by the CS due to the externally modulated pump. We also extend the \textcolor{black}{variational} approach to solve two solitons interaction problem in the absence as well as in the presence of the externally modulated field. In absence of phase modulated field, the two co-propagating solitons can attract, repulse or \textcolor{black}{can propagate independently depending on their initial delay.} The final state of interaction can be predicted through a second-order differential equation which is derived by the variational method. While in presence of the phase modulated field, the two solitons interaction can result in annihilation, merging, breathing or \textcolor{black}{two soliton} state depending on the detuning frequency and the pump power.  Variational treatment analytically predicts these states and portrays the related dynamics that agrees with full numerical simulation carried out by solving the normalized Lugiato-Lefever equation. The results obtained through this \textcolor{black}{variational approach} will enrich the understanding of complex pulse dynamics under phase modulated driving field in passive dissipative systems.

\end{abstract}

\pacs{Valid PACS appear here}
\maketitle


\section{\label{sec:level1}INTRODUCTION}

Temporal cavity solitons (CSs) are the localized pulse that coexists with a homogeneous continuous wave (CW) background observed in driven nonlinear passive resonators \cite{cs1,cs2,cs3}. Unlike the conservative system where nonlinearity balances the pulse broadening due to chromatic dispersion \cite{inbook}, solitons in the dissipative system need an extra balance of energy to compensate for the total loss \cite{in1book}. The total loss in these dissipative systems is either balanced by an external CW pump or by the presence of an active gain medium. CS was first observed in 2010 in a single mode fiber ring resonator \cite{fo} and since then it has gained considerable attention towards all optical buffers \cite{of1,of2}. They have also been observed in Kerr microresonators, which enables on-chip frequency comb generations \cite{fc1,fc2}. Soliton propagation within a resonator under phase modulated driving field has  obtained numerous attention \cite{pm1,pm2,pm3,doi:10.1063/1.2828458,doi:10.1063/1.2388867} because with the minimum modulation depth one can obtain a deterministic way of generating  CS without undergoing a chaotic phase. Moreover, it helps to achieve greater control over the generated CS. On the other hand, a theoretical tool depending on the variational approach was first introduced by Anderson \cite{vc1} for nonlinear pulse propagation, since then it has been used extensively in conservative systems \cite{vc2,vc3,var_dis2} as well as in dissipative systems including active \cite{vd1,vd2} and  passive resonators \cite{var-dissipative,bs2,var_dis1} to study the pulse dynamics. In the first part of our study, we made a detailed analysis of the dynamics of a single CS in the framework of variational treatment and  obtained several closed form expressions of the CS characteristics which \textcolor{black}{can be determined} according to the phase profile of the CW. To illustrate this perturbation problem, we numerically solved the Lugiato-Lefever equation (LLE) \cite{l1,l2,PhysRevA.82.033801,PhysRevA.89.063814} and compared the results obtained with the variational approach.
\par CSs are generally very robust in nature. One of the key features of CSs is that they can be individually addressed without affecting their nearest neighbour soliton and several of them can even co-exist and propagate independently \cite{cs2,vd2,article-mul}. Thus an important limitation may occur if they interfere with each other while propagating. It will disrupt the information of a bit pattern after few round trips within the cavity. \textcolor{black}{Recently bound state of two solitons in dissipative passive resonator system has been studied \cite{bs3,bs4}, where the stable and unstable equilibrium separation between CS has been attributed by calculating the maxima and minima of their interaction potential.} In another article \cite{ip1}, the controlled interaction between two CSs are studied  where merging and annihilation of two solitons  are observed against suitable driving strength and frequency detuning. Here, we consider interaction picture of two CSs both in absence and in presence of a phase-modulated driving field in the framework of variational analysis with the addition of a Rayleigh's Dissipation function (RDF) \cite{rdf1} that takes into account the perturbations within the system . Though the phenomena of single CS dynamics in such a system is well understood in presence of phase modulated driving field, the analytical closed form expressions for different properties has not been investigated earlier. The closed form analytical expressions for dynamic pulse parameters are useful in realizing the underpinning physics. Also, our findings would extend the importance of variational method as a powerful theoretical tool which can handle multisolitons dynamics and even identify the individual effects of internal perturbation (soliton interactions) and external perturbations (phase or amplitude modulated driving field). Thus, we believe our findings with this semi-analytical technique will help further to deal with several real perturbation situations in the passive dissipative resonator systems complementing various experimental and numerical results.\par
We organize our work as follows : In Sec. II, we model the localized CS by mean-field LLE and briefly introduce the \textcolor{black}{variational technique}. In Sec. III, we discuss the results of a single soliton dynamics under a phase modulated CW driving field. We assume a cosine modulated phase profile of the input laser which can be created with an electro-optic modulator and the modulation frequency to be identical with the cavity FSR. Numerical simulations for the dynamics of a single CS are performed with normalized modified LLE incorporating the phase-modulated pump. The variational technique further describe the results analytically and unfold the underlying physics. In Sec. IV, we consider the two-soliton interaction dynamics \textcolor{black}{due to constant driving field} with the help of this semi-analytical technique. In Sec. V, we investigate various complex dynamics incorporating the above mentioned two perturbations simultaneously with a fixed delay between the two CSs i.e. two co-propagating CS under pump phase-modulated driving field. \textcolor{black}{Stable two soliton, merged, annihilated and breathing single soliton states} are formed with  suitable choice of detuning frequency and pump power.  

\section{General Model}
\subsection{\textcolor{black}{Mean-field approach}}
We consider an optical fiber loop resonator that exhibits Kerr nonlinearity with anomalous dispersion. The evolution of the slowly varying intra-cavity field envelope $\psi(t,\tau)$ is modeled by the following  mean-field LLE \cite{l1,cs3}:
\begin{equation}
\tau_R\frac{\partial\psi}{\partial \tau}=-(\alpha+i\nu)\psi+i\gamma L_c|\psi|^2\psi-i\frac{\beta_2 L_c}{2}\frac{\partial^2\psi}{\partial t^2}+\sqrt{\theta}E_{in}.
\label{eq.1}
\end{equation}

\begin{figure}[htbp]
	\centering
	\includegraphics[width=\linewidth]{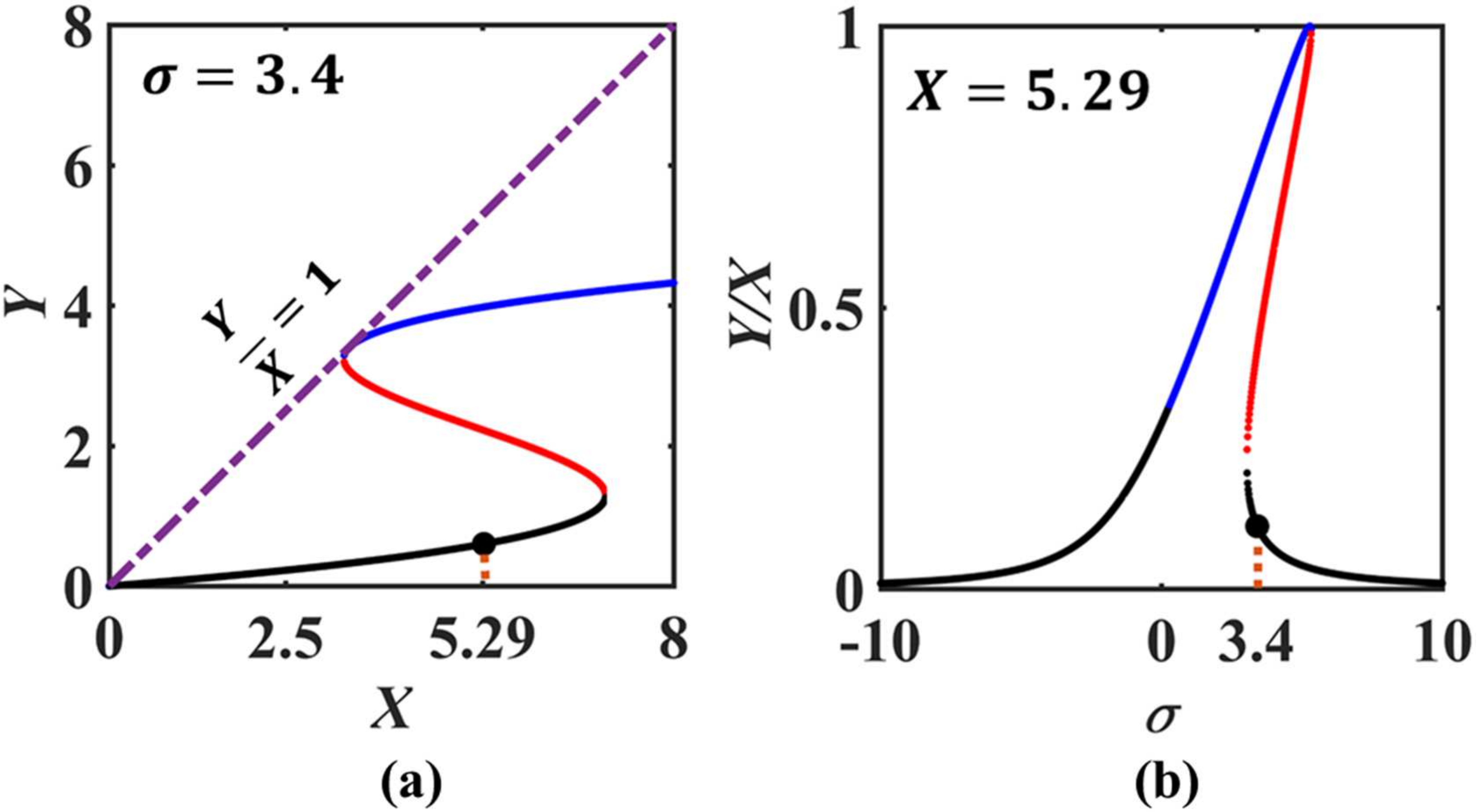}
	\caption{Optical bistability of the steady and homogeneous state of intra-cavity field. (a) Bistability curve showing intra-cavity power $(Y)$ vs. normalized pump power $X$; black circle indicates the chosen pump power for simulation (b) Ratio of the intra-cavity power $(Y)$ to pump power $(X)$ vs. normalized detuning $(\sigma)$; tilted under Kerr nonlinearity, black circle indicates the operating detuning frequency.} 
	\label{fig:1}
\end{figure}

Here, $\tau$ is the normalized slow time scale related with the total evolution time within the resonator, and $t$ denotes the fast time which characterize the temporal envelope profile of the generated CS in a reference frame moving at the group velocity of the external driving field, $\tau_R$ is the round trip time within the cavity. These three time scales can be linked to the round trip index number $m$ as $E(t=mt_R,\tau)=E^{(m)}(0,\tau)$. The terms in the right hand side of Eq.(\ref{eq.1}) signify losses $(\alpha)$, nonlinear phase detuning $(\nu)$ of the driving field $E_{in}(t,\tau)$ from the nearest cavity resonance, $\gamma$ is the fiber nonlinear coefficient, $\beta_2<0$ signifies second order anomalous dispersion and $E_{in}$ is the external coherent CW driving field , $\theta$ is the power transmission coefficient from coupler to the resonator and the length of the resonator is $L_c$. Introducing the normalization factors : $\tau$ $\,\to\,$ ${\alpha \tau}/{\tau_R}$, $t$ $\,\to\,$ $t({2\alpha}/{|\beta_2|L_c})^{{1}/{2}}$, and the intra-cavity field $\psi$ $\,\to\,$ $\psi(\gamma L_c/\alpha)^{{1}/{2}}$, normalized detuning frequency $\sigma = \nu/\alpha$, normalized pump $E_{in}$$\,\to\,$$E_{in}(\gamma L_c\theta/\alpha^3)^{{1}/{2}}$ , Eq.(\ref{eq.1}) takes the following form:

\begin{equation}
\frac{\partial \psi}{\partial \tau}=-(1+i\sigma)\psi+i|\psi|^2\psi+i\frac{\partial^{2}\psi}{\partial t^2}+E_{in}.
\label{eq.2}
\end{equation}
\par The steady-state and homogeneous solution of Eq.(\ref{eq.2}) satisfies the well-known cubic steady-state equation \cite{HAELTERMAN1992401} $X=Y^3-2\sigma Y^2+(\sigma^2+1)Y$  with $X=|E_{in}|^2$ and $Y=|\psi|^2$. The steady state curve is single valued for $\sigma<\sqrt{3}$ and the curve takes an $S$- shape in case of $\sigma>\sqrt{3}$. It has been calculated in \cite{HAELTERMAN1992401} that for $\sigma \geq 2$, the power threshold for modulation instability (MI) to occur in anomalous dispersion regime requires $Y>\sigma/2$, which makes the entire upper branch of Fig.\ref{fig:1}(a) unstable, so we choose the operating power in the lower stable branch of bistability curve. Fig.\ref{fig:1}(b) can be interpreted as the resonance
of the ring cavity which is tilted in presence of the Kerr nonlinearity where we have marked (filled circle) the operating detuning frequency $(\sigma)$ in this figure.   

\subsection{\textcolor{black}{Variational approach}}
We introduce an analytical model which can support the numerical simulation results obtained via LLE. To analyze a complex situation like soliton interaction in presence of an external perturbation, we deal two problems individually with the help of  variational technique. It helps us to gain more insight about the physical parameters of CS that undergo changes in presence of phase modulated driving field and also due to soliton interactions. The formalism of the variational technique is based on the proper choice of an ansatz function. Though a  functional form of soliton pulse in dissipative system is presented in \cite{PhysRevE.62.8726}, we found a rather easy \textit{sech} ansatz is also suitable to start with \textcolor{black}{in such dissipative systems which has previously used in \cite{cs3,bs2,var_dis1}.} Further, we introduce the RDF \cite{rdf1} to handle the perturbations. The construction of RDF has been performed in such a way that the generalized Euler-Lagrange (EL) equation must reproduce the LLE and its complex conjugate. \textcolor{black}{First,} we reduce the problem by inserting the ansatz into Lagrangian and RDF function. \textcolor{black}{Then} integrating it over fast time $t$ we get : $L_g=\int_{-\infty}^{\infty} L\hspace{1 mm}dt$, and
$R_g=\int_{-\infty}^{\infty} R\hspace{1 mm}dt$. For different perturbation situations  the reduced Lagrangian  $(L_g)$ remains same,  where the reduced RDF  $(R_g)$ changes according to the perturbations faced by  CS. Finally, we exploit the EL equation to obtain the equation of motion of different pulse parameters:

\begin{equation}
\frac{d}{d\tau}\left(\frac{\partial L_g}{\partial \dot{p_j}}\right)-\frac{\partial L_g}{\partial p_{j}}+\left(\frac{\partial R_g}{\partial \dot{p_j}}\right) = 0,
\label{eq.3}
\end{equation}
here $p_j$ stands for different pulse parameters (like amplitude, position, phase , frequency etc.) and $\dot{p_j}=\frac{dp_j}{d\tau}$.

\section{Soliton Propagation in a Cavity with Phase-modulated Driving Field }
\label{3}
We consider a situation when the phase of a driving beam is modulated with a cosine profile instead of a constant value and modulation frequency is identical with the resonator free spectral range. Data for numerical simulation is identical to the one reported in \cite{ip1}. The driving field $E_{in}$ of Eq.(\ref{eq.2}) becomes a function of fast time $(t)$, we assume it of the form,
\begin{equation}
E_{in}(t)= P_{in}\exp(i M\chi(t)),
\label{eq.4}
\end{equation}
where, $\chi(t)=\cos(\omega t)$,  $M$ is the modulation depth and $\omega$ is the modulation frequency. The driving CW field has power $P_{in}$ where it's phase is modulated with the help of an electro-optic modulator \cite{pm1}. For numerical simulations, we use normalized parameters, $\sigma = 3.4$, $P_{in}=2.3$, $M=0.15$, $\omega=0.05$ and we choose our initial wave function of the form : $\psi(0,t)=\sqrt{2\sigma}sech(\sqrt{\sigma}(t-t_p))$ ; $t_p = 5$. Under phase-modulation, CS undergoes a temporal shift  within the resonator. 
\subsection{\textcolor{black}{Derivation of the reduced model}}
 To gain more insight, we develop a perturbative theory based on variational method. We proceed our analysis with the standard Kerr soliton ansatz \cite{rdf1}:
 \begin{figure}[tp]
	\centering

	\includegraphics[trim=0.0in 0.0in 0.0in 0.0in,clip=true,  width=60mm]{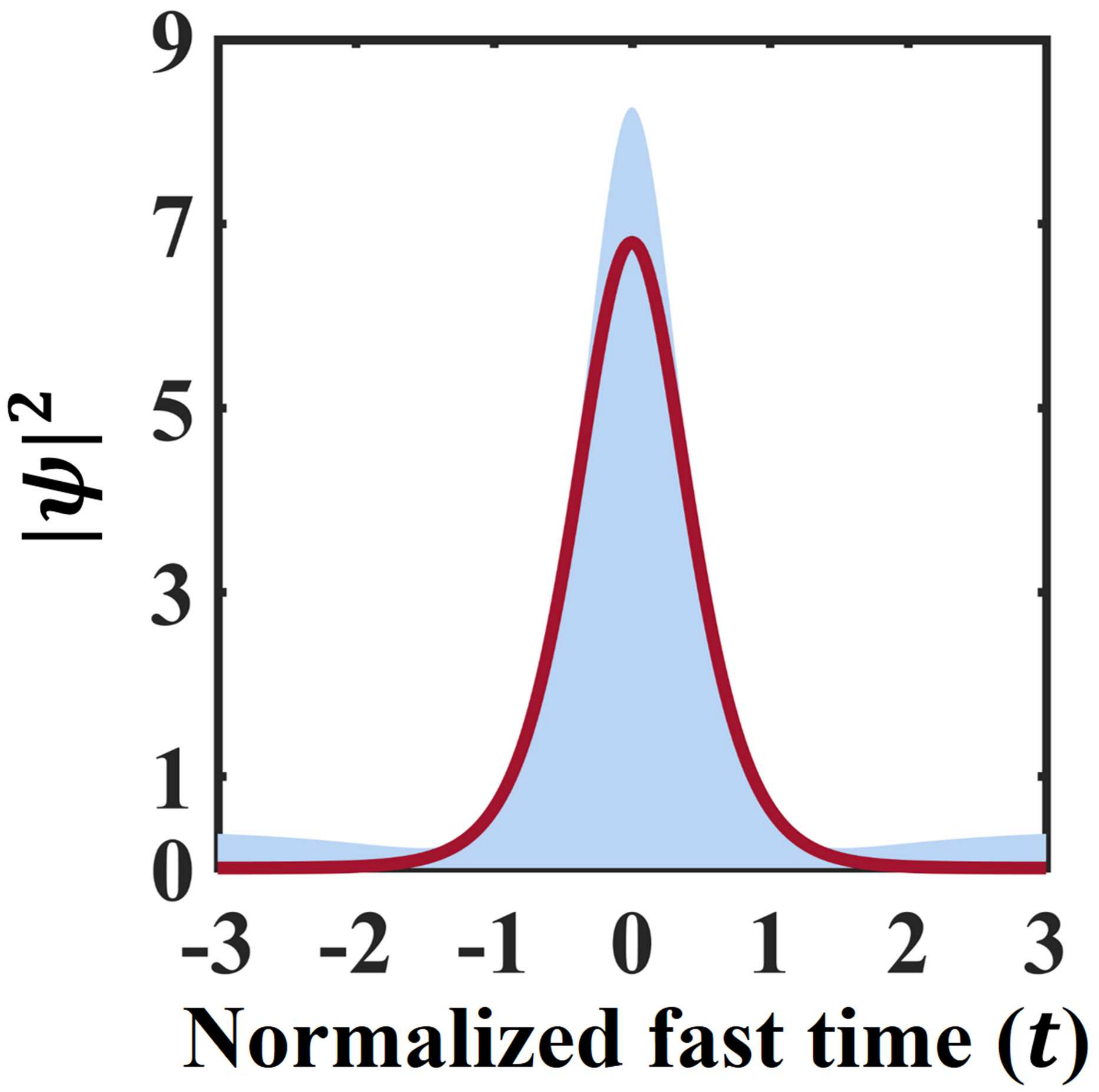}
	\caption{Comparison of the actual CS (shaded) and the ansatz (solid red line) used in variational formulation.}
	\label{fig:2}
\end{figure}
\begin{multline}
\psi(t,\tau)=\sqrt{2\eta(\tau)}sech(\sqrt{\eta(\tau)}(t-t_p(\tau)))\\\times \exp[i(\phi(\tau)-\delta(\tau)(t-t_p(\tau)))],
\label{eq.5}
\end{multline}
where the amplitude \textcolor{black}{$(\sqrt{2\eta})$}, phase $(\phi)$, position $(t_p)$ and frequency shift $(\delta)$  are allowed to vary over slow time. \textcolor{black}{Note, the accuracy of the variational method depends on the proper choice of the ansatz. Unlike Schr\"{o}dinger Kerr soliton, the CS is formed over a homogeneous pedestal. The given $sech$ ansatz doesn't include this pedestal and slightly misfit from the actual CS as depicted in Fig \ref{fig:2}. Note that, in the formalism of the variational method we require an ansatz which is square integrable (i.e
$\int_{-\infty}^{\infty} |\psi|^2 dt<\infty$). It is difficult to form the reduced Lagrangian  $L_g (=\int_{-\infty}^{\infty}L dt)$ if the given ansatz contain a constant pedestal.  Hence lack of pedestal in the $sech$ form may leads to an error in peak power calculation. However, this error can be reduced by eliminating the homogeneous background through the resealing of field   power as, $|\psi(t,\tau)|^2 \to |\psi(t,\tau)|^2-|\psi_h|^2$, where $\psi_h\approx E_{in}\sigma^{-1}(\sigma^{-1}-i)$ is the homogeneous field describing the CW background \cite{cs3}.} Next we construct the Lagrangian $(L)$ and RDF $(R)$  for LLE given in Eq.(\ref{eq.2}) as follows:
\begin{equation}
L= \frac{i}{2}(\psi\psi_\tau^*-\psi_\tau\psi^*)+|\psi_t|^2-\frac{1}{2}|\psi|^4+\sigma|\psi|^2,
\label{eq.6}
\end{equation}

\begin{equation}
R=i(\psi\psi_\tau^*-\psi_\tau\psi^*)+i(E_{in}^{*}\psi_\tau-E_{in}\psi_\tau^*).
\label{eq.7}
\end{equation}
\par It is observed that even a small value of modulation depth  $(M<<1)$ can influence the trajectory of the CS. Therefore, we approximate $E_{in}$ \textcolor{black}{Eq.(\ref{eq.4})} as:
\begin{multline}
E_{in}(t) \approx P_{in}(1+i M \cos(\omega t))
\\\approx P_{in}\left(1+i M-\frac{i M \omega^2 t^2}{2}\right).   
\label{eq.8}
\end{multline}
\begin{figure}[tp]
	\centering
	\includegraphics[width=\linewidth]{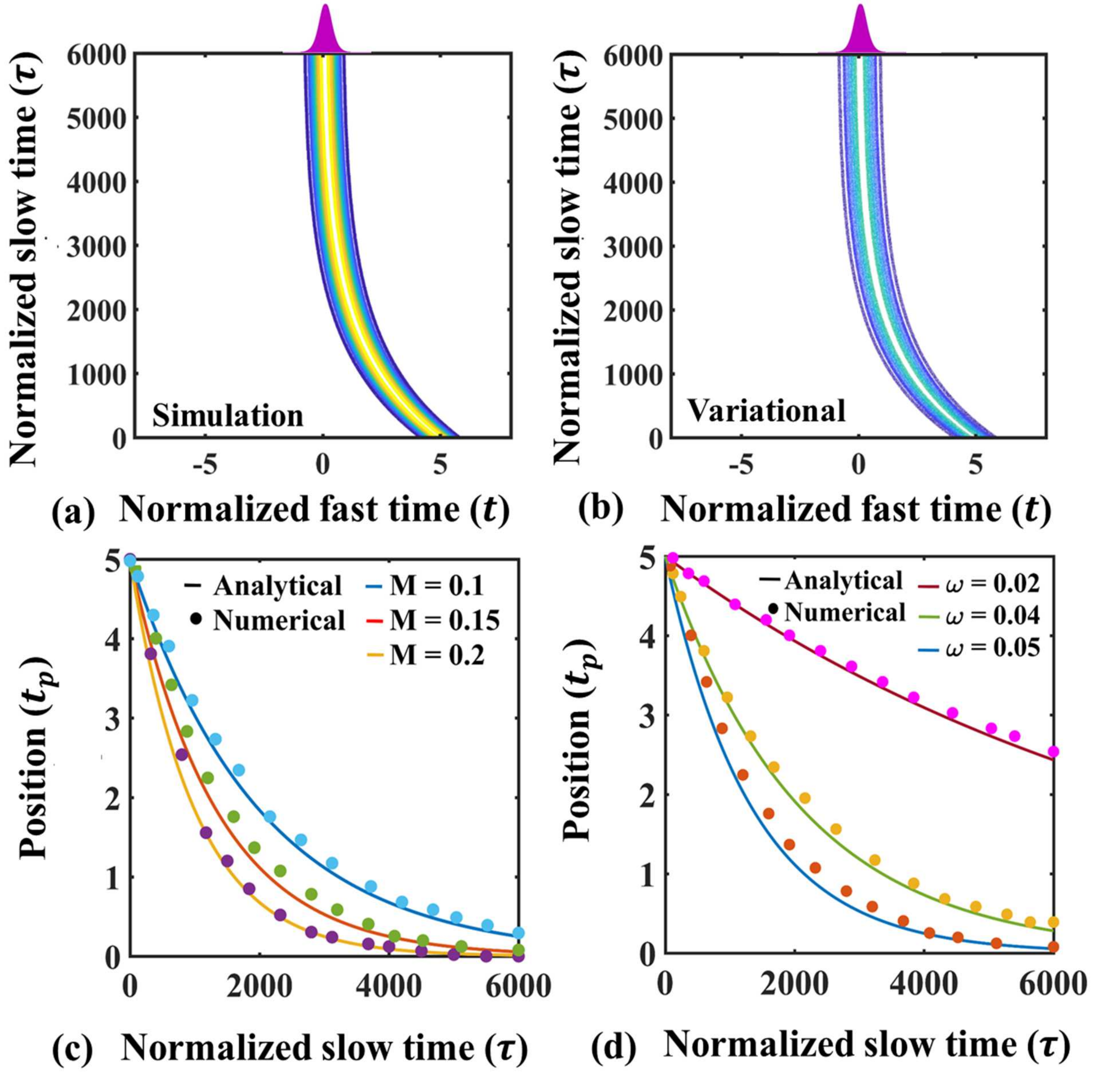}
	\caption{Contour plots of (a) Simulation (b) Variational results showing the trajectory of CS under the influence of phase-modulated driving field. (c) Variation of CS trajectory with different phase modulation depth $(M)$ and (d) Different phase modulation frequency $(\omega)$ of CW, shown both numerically and analytically.} 
	\label{fig:3}
\end{figure}
\par Exploiting the EL equation with the help of reduced form of Lagrangian $( L_g=\int_{-\infty}^{\infty} L\hspace{1 mm}dt)$ and RDF $(R_g=\int_{-\infty}^{\infty} R\hspace{1 mm}dt)$ (see derived form of $L_g$ and $R_g$ in appendix A), we derive a set of coupled ordinary differential equations (ODEs) describing the overall soliton dynamics ($\eta,\delta,\phi, t_p$). The equations of motion for individual pulse parameters are as follows:
\textcolor{black}{
\begin{multline}
\frac{\partial\eta}{\partial\tau}=-4\eta+ \frac{\sqrt{\eta}}{2}\Bigl(2\sqrt{2}\pi P_{in}-\frac{\pi^3\delta^2 P_{in}}{2\sqrt{2}\eta}\Bigr)\Bigl(\cos\phi+M\times\\ \sin\phi\Bigr)-\frac{M\omega^2\sqrt{\eta}}{4}\times\Biggr(2\sqrt{2}P_{in}\sin\phi\left(\frac{\pi^3}{4\eta}+\pi t^2_p\right)-\frac{2\delta^2 P_{in} }{\eta}\\ \times \sin\phi\left(\frac{\pi^3 t_0^2}{4}+\frac{5 \pi^5}{16\eta}\right)-\frac{\sqrt{2}\delta \pi^3 P_{in} t_p \cos\phi}{\eta}+\frac{5\sqrt{2}\pi^5\delta^3 P_{in} }{24\eta^2}\\ \times t_p \cos\phi\Biggr)
\label{eq.9}
\end{multline}
\begin{multline}
\frac{\partial\phi}{\partial\tau}=-\sigma+\eta+\delta^2+\frac{\pi^3P_{in} \delta^2}{2\sqrt{2}\eta^{3/2}}\Biggl(\sin\phi +M \cos\phi\Biggr) +\\\frac{5\sqrt{2}M\omega^2\pi^3\delta P_{in}t_p \sin\phi}{8\eta^{3/2}}-\frac{5\pi^5\delta^4 P_{in} M \cos\phi}{96\sqrt{2}\eta^{3/2}}-\\\frac{55\sqrt{2}\delta^2 P_{in} t_p \pi^5 M\omega^2 \sin\phi}{192\eta^{3/2}}-\frac{M\omega^2 P_{in}\delta^4 \cos\phi}{12\sqrt{2}\eta^{3/2}}\Biggl(\frac{61\pi^7}{64\eta}\\+\frac{5\pi^5 t_p^2}{16}\Biggr)-\frac{M \omega^2 P_{in} \delta^2 \cos\phi \pi^3}{8\eta^{3/2}}\Biggl(\frac{25\sqrt{2}\pi^2}{8\eta}+\frac{3t_p^2}{\sqrt{2}}\Biggr)
\label{eq.10}
\end{multline}
\begin{multline}
\frac{\partial\delta}{\partial\tau}=-\frac{\delta}{4\sqrt{\eta}}\Bigl(2\sqrt{2}\pi P_{in}-\frac{\pi^3\delta^2 P_{in}}{2\sqrt{2}\eta}\Bigr)\Bigl(\cos\phi+M \sin\phi\Bigr)\\+ \frac{M\omega^2}{4\sqrt{\eta}}\times\Biggl(2\sqrt{2}\pi P_{in}t_p  \cos\phi+\sqrt{2}\delta P_{in}\pi t_p^2 \sin\phi\times\\ \Bigl(1- \frac{\delta^2\pi^2}{4\sqrt{2}\eta}\Bigr)+\frac{3\delta P_{in}\pi^3 \sin\phi}{2\sqrt{2}\eta}-\frac{\sqrt{2}\pi^3\delta^2 P_{in}t_p \cos\phi}{\eta}\times \\\Bigl(1+\frac{5\pi^2}{96\eta}\Bigr)\Biggr)
\label{eq.11}
\end{multline}
\begin{multline}
\frac{\partial t_p}{\partial\tau}=-2\delta+\frac{\delta P_{in}\pi^3}{4\sqrt{2}\eta^{3/2}}\Bigl(\sin\phi-M \cos\phi\Bigr)-\frac{M\omega^2}{4\sqrt{2\eta^{3}}}\times\\ \Biggl(P_{in}t_p\pi^3 \sin\phi-2\sqrt{2}\delta P_{in}\cos\phi \Biggl(\frac{5\pi^5}{16\eta}+\frac{\pi^3t_p^2}{4}\Biggr)\\-\frac{5\pi^5\delta^2 P_{in}t_p \sin\phi}{8\sqrt{2}\eta}+\frac{P_{in} \delta^3 \cos\phi}{3\sqrt{2}\eta}\Biggl(\frac{61\pi^7}{64\eta}+\frac{5\pi^5t_p^2}{16}\Biggr)\Biggr)
\label{eq.12}
\end{multline}}
\par \textcolor{black}{
 To obtain closed form expressions for CS trajectory as well as the peak amplitude and phase of the generated CS in the steady state, we found the contributions of those terms containing higher orders of $\delta$, ($\delta^n,  n \geq 2$) in Eqs.(\ref{eq.9})-(\ref{eq.12}) are negligible.}

\subsection{\textcolor{black}{Prediction of the soliton trajectory}}
\begin{figure}[tp]
	\centering
	\includegraphics[width=\linewidth]{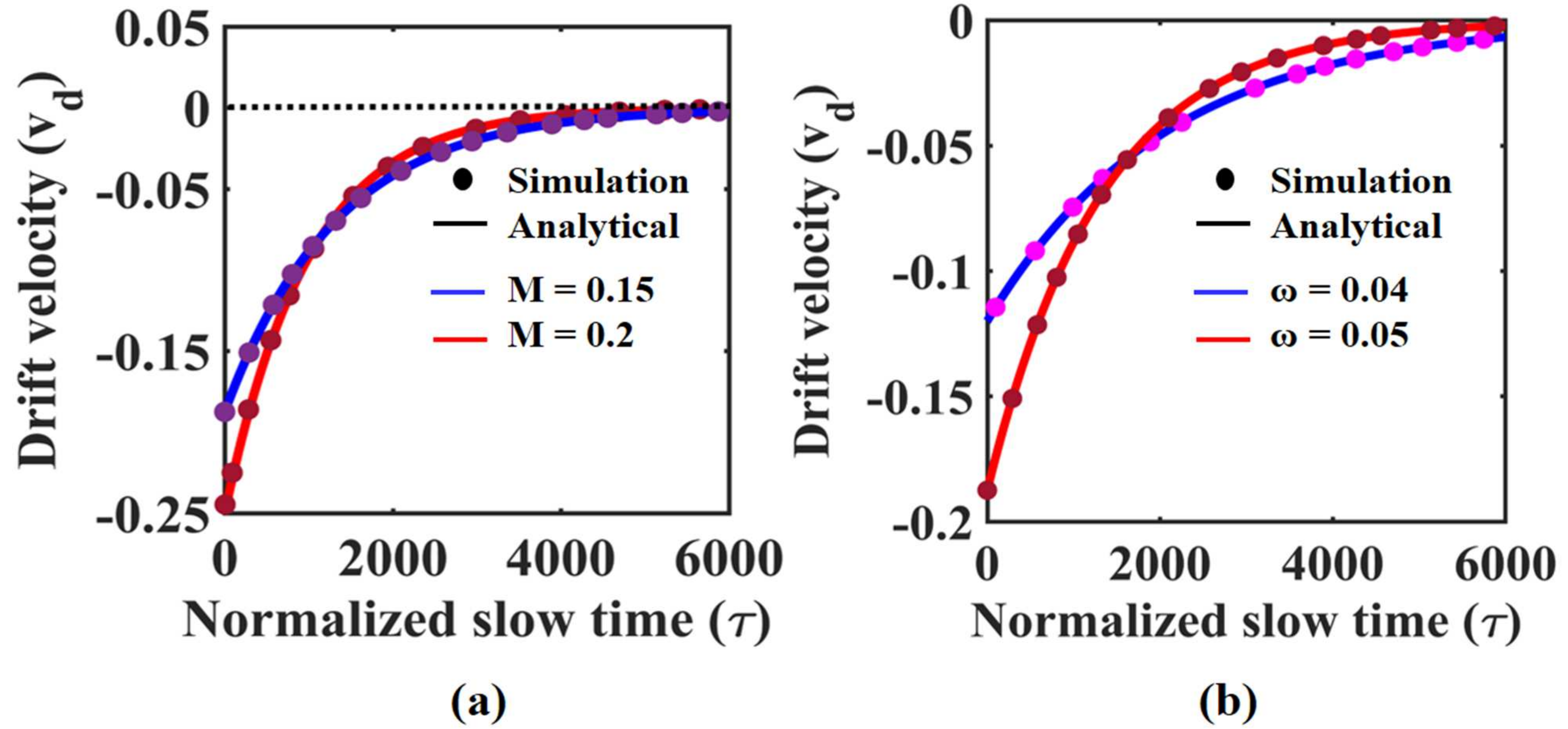}
	\caption{Drift velocity profile of the CS shown both numerically and analytically (a) CS drift velocity vs. normalized slow time for two modulation depth $(M = 0.15, M =0.2)$ (b) CS drift velocity vs. normalized slow time for two modulation frequency $(\omega = 0.04, \omega =0.05)$. }
	\label{fig.4}
\end{figure}

From \textcolor{black}{the mathematical expressions Eqs.(\ref{eq.9})-(\ref{eq.12})}, it is evident that pulse parameters undergo some changes during propagation for non-vanishing $M$. In Fig.\ref{fig:3}(a), we plot the trajectory of a CS under phase-modulated external pump.  We observe that, CS acquires a drift velocity during its propagation. The trajectory of the soliton is influenced by the modulation depth. The numerical solution of the coupled ODEs predict the trajectory nicely (see Fig.~\ref{fig:3}(b)). However, those ODEs will be more useful if we decouple them with suitable approximations \textcolor{black}{( see appendix B).} It can be shown that the dynamic expression for temporal shift of the pulse is:
\begin{equation}
t_p(\tau)\approx t_p(0)\exp(-2M\omega^2\tau)
\label{eq.13}
\end{equation}
Eq.(\ref{eq.13}) clearly indicates with increase in slow time $(\tau)$,  the location of the pulse shifts towards origin ($t=0$). Here modulation depth $(M)$ and modulation frequency $(\omega)$ act as parameters and control the drift velocity of the generated CS. The pulse shifts to its stationary location at less round trips with increasing $M$ or $\omega$ (see Fig.~\ref{fig:3}(c) and \ref{fig:3}(d)). The drift velocity has been previously mentioned to be directly proportional to the gradient of external driving field \cite{PhysRevE.62.8726,l2}. Our calculations enable us to obtain a simple mathematical form for the drift velocity of generated CS, which is given as $v_d(\tau) = \frac{d t_p}{d \tau} \approx -2M\omega^2 t_p(\tau)$. In Fig.(\ref{fig.4}) we observe that drift velocity of the CS initially increases and gradually saturates to null value and thus CS obtains its stationary position.

\begin{figure}[bp]
	\centering
	\includegraphics[width=\linewidth]{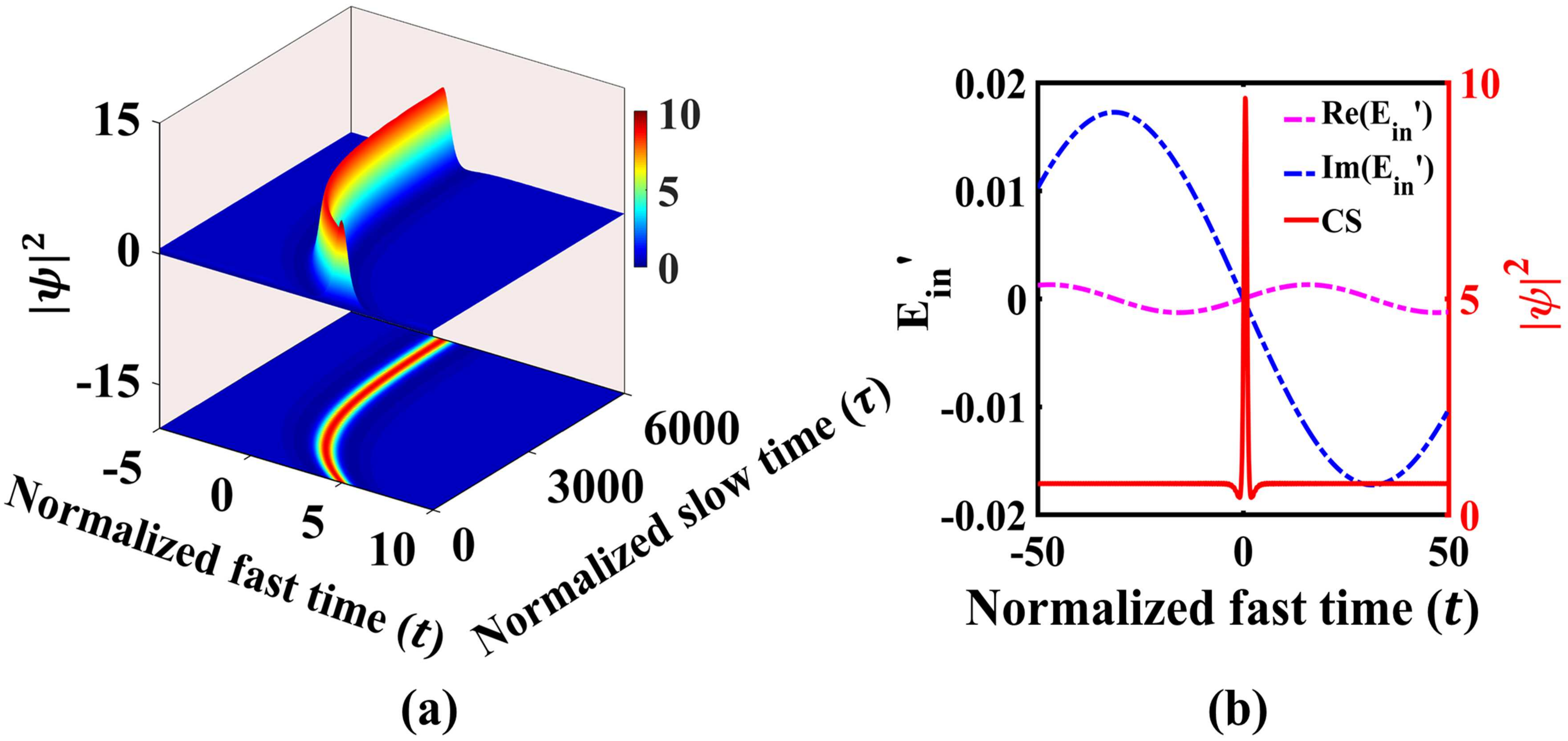}
	\caption{(a) Steady position of CS undergoes a shift in its reference frame due to phase modulated CW pump ($E_{in}=P_{in}e^{\phi(t)}$). (b) (Left axis) Gradient of external pump field $\frac{d E_{in}}{d t}= E_{in}^{'}$ ; (Right axis) Final steady position of generated CS.}
	\label{fig.5}
\end{figure}

\par The behavior of the CS trajectory can be understood in an alternative way with the concept of intra-cavity field momentum, which is defined by \cite{pm1}:

\begin{equation}
P=-\frac{i}{2} \int_{-\infty}^{\infty} (\psi^{*}\frac{d\psi}{dt}-\psi\frac{d\psi^{*}}{dt})\hspace{1 mm}dt.
\label{eq.14}
\end{equation}
Using this definition along with Eq.(\ref{eq.2}), the rate of change of momentum of intra-cavity field is found to be,
\begin{equation}
\frac{dP}{d\tau}=-2P- Im\int_{-\infty}^{\infty} \psi\frac{d E^{*}_{in}}{dt}\hspace{1 mm}dt.
\label{eq.15}
\end{equation}
\begin{figure}[tp]
	\centering
	\includegraphics[trim=0.3in 0.1in 0.1in 0.2 in,clip=true,  width=85mm]{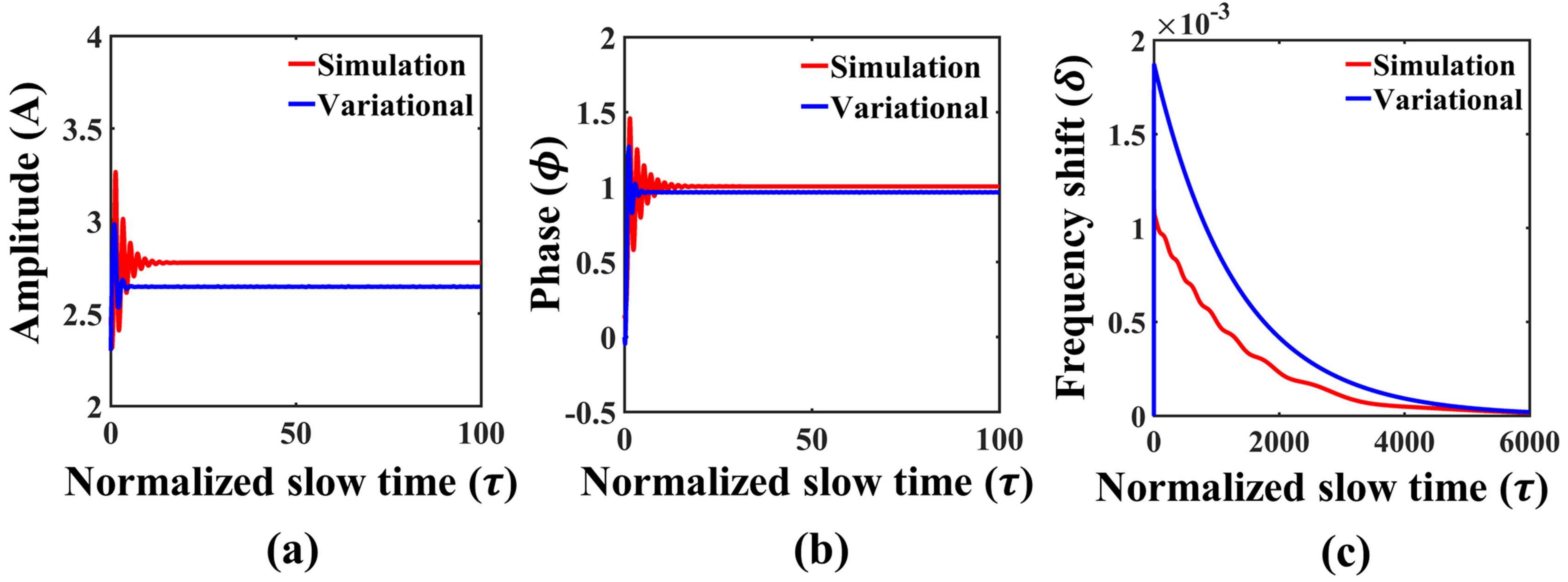}
	\caption{Evolution of CS parameters under phase modulated driving field shown both numerically and with variational analysis (a) Amplitude (b) Phase and (c) Frequency shift. }
	\label{fig.6}
\end{figure}
\par In absence of any modulation (phase or amplitude) of the pump, the rate of change of momentum is determined only by the first term of the right hand side of Eq.(\ref{eq.15}) and the momentum seems to decay exponentially with $\tau \rightarrow \infty$  such that the net force acting on the soliton also vanishes with propagation and the soliton eventually acquire its steady position. \textcolor{black}{Note that in this problem the phase of the driving field is modulated as, $\phi(t)=M\chi(t)=M \cos(\omega t)$.} In presence of phase modulation (i.e $E_{in}=P_{in}e^{\phi(t)}$) the second term of the Eq.(\ref{eq.15}), which contains the gradient of the driving field, will dominate. \textcolor{black}{Now, CS will gradually acquire its stationary position where the gradient of the driving field i.e. $dE_{in}/dt$ (or $\frac{d \phi}{dt}$ ) vanishes. Note that, $\psi$ and $E_{in}$ both are in general complex with a phase term. since the amplitude of $\psi$ (which represents soliton) is a symmetric function it is easy to show that the integrand in Eq. (\ref{eq.15}) vanishes when $\frac{d \phi}{dt}$ is an asymmetric function or zero. Now when the function $\frac{d \phi}{dt}$ is neither symmetric nor asymmetric the integrand can still vanishes, if $\frac{d \phi}{dt}=0$ and the soliton is dragged to that point. We illustrate this phenomenon in  Fig. \ref{fig.5} where soliton is dragged at the point where $dE_{in}/dt$ (or $\frac{d \phi}{dt}$ ) vanishes.} Note that, the concept of the intra-cavity field momentum only locates the stabilized temporal position of an off-shifted CS under phase-modulated pump. It can never predict the pulse trajectory like variational treatment does. The variational treatment can be extended further to evaluate the other pulse parameters like amplitude $(A=\sqrt{2\eta})$, phase $(\phi)$ and frequency shift $(\delta)$. In Fig.(\ref{fig.6}) we plot the variation of $A$, $\phi$ and $\delta$ as a function of slow time ($\tau$). \textcolor{black}{We know, the dissipative CS always emerge out of the CW background unlike the Kerr solitons in the conservative system. During our variational analysis we assume a $sech$ pulse shape which is merely an approximation. Note, in order to get rid of the homogeneous pedestal ($\psi_h$), we rescale the field as, $|\psi(t,\tau)|^2 \to |\psi(t,\tau)|^2-|\psi_h|^2$ and compare it with variational result in Fig.(\ref{fig.6}).  It is observed that the variational results are in good agreement with the results obtained from full numerical solution of LLE.} 

\subsection{\textcolor{black}{Derivation of the stationary solution}} By exploiting the variational expressions we extract the steady-state condition for CS. Without any phase-modulation $(M=0)$, from Eqs.(\ref{eq.9})-(\ref{eq.10}), we obtain the amplitude and phase of CS in its steady state as:

\begin{equation}
\begin{cases}
\eta = \sigma \\
\phi = \cos^{-1}\left(\frac{2\sqrt{2\eta}}{\pi P_{in}}\right) = \cos^{-1}\left(\frac{2\sqrt{2\sigma}}{\pi P_{in}}\right).
\end{cases}
\label{eq.16}
\end{equation}

\begin{figure}[bp]
	\centering
	\includegraphics[width=\linewidth]{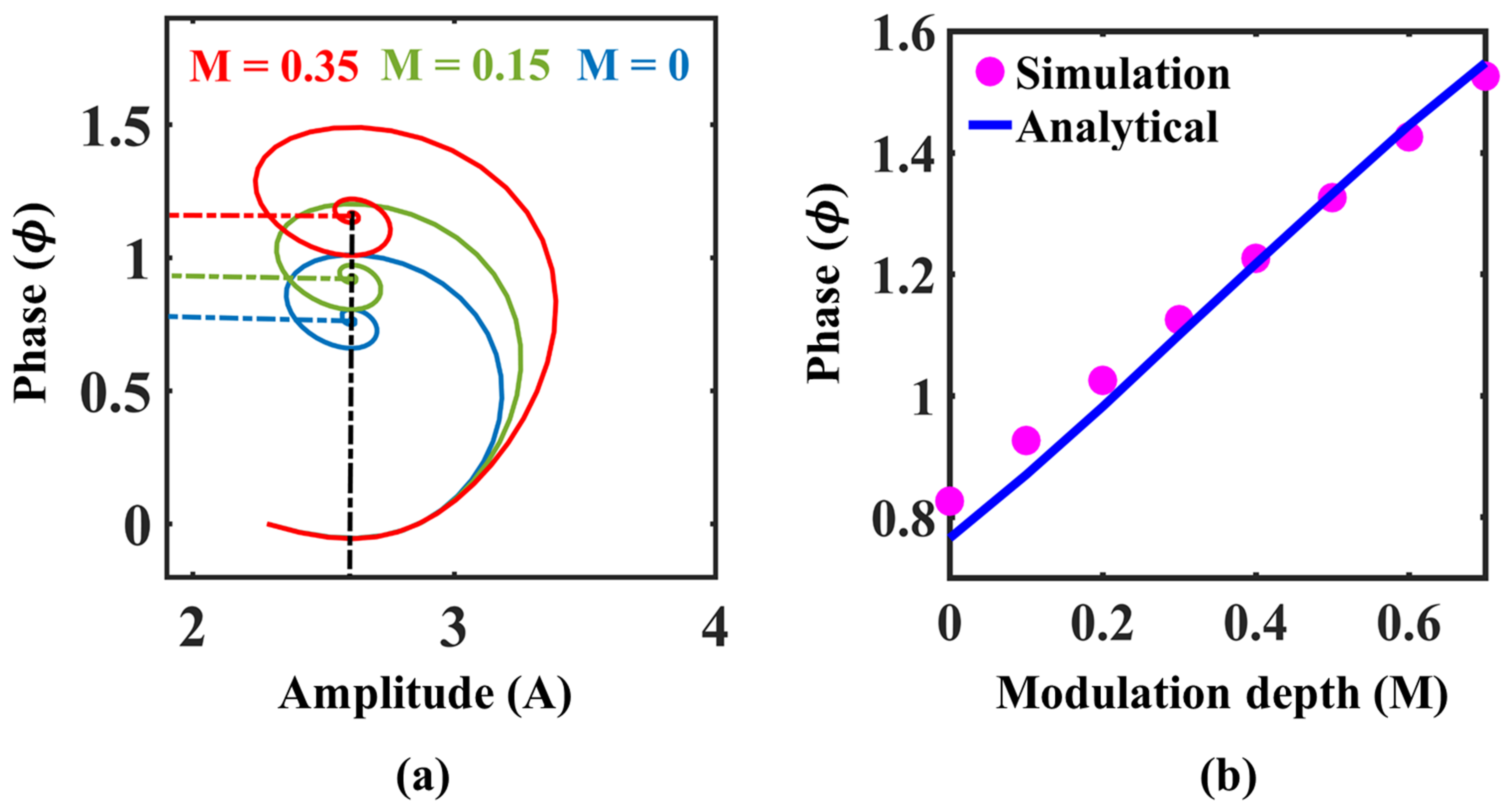}
	\caption{(a) Phase-space diagram of generated CS in three different cases, without modulation ($M$), $M=0.15$, $M=0.35$ respectively. (b)  Variation of soliton phase with modulation depth $(M)$ ; shown for both numerical and analytical results.}
	\label{fig.7}
\end{figure}

\par From  Eq.(\ref{eq.16}) it is evident that, for a given pump power ($P_{in}$) the CS ceases to exist when $\sigma > \frac{\pi^2 P_{in}^2}{8}$.  We numerically verify this argument by keeping $P_{in} = 2.3$ and find that CS is generated when $\sigma\leq 6.5$, which is consistent with the expression we have derived. In a phase-space diagram, CS always makes a spiral \cite{pm2} centered about a point known as fixed point of the system. This spiral nature is a signature of the evolution of amplitude and phase of the generated CS.  Under phase-modulation, the fixed point might change as depicted in Fig.\ref{fig.7}(a). For non-vanishing $M$, Eq.(\ref{eq.16}) is modified as,  

\begin{equation}
\begin{cases}
\eta \approx \sigma \\
\phi \approx tan^{-1} M+\cos^{-1}\left(\frac{B}{\sqrt{1+M^2}}\right),
\end{cases}
\label{eq.17}
\end{equation}
\\where, $B = \frac{2\sqrt{2\sigma}}{\pi P_{in}}$. In Fig.\ref {fig.7}(b), we plot CS phase $(\phi)$ as a function of modulation depth $(M)$. We observe that variational results (solid line) agree well with full numerical results (filled circles).   

\section{Soliton interaction in the presence of a constant driving field}
\label{4}

Earlier work of Malomed and Akhmediev and later several other authors have considered the detailed analysis of soliton (bright-bright, bright-dark) interaction  problem in conservative system \cite{article_pp,bs1,bs5,PhysRevE.47.2874,article00,Smith:94,Maruta:95}. Stability of two and multi-soliton states has also been reported for dissipative solitons where the system contains a gain medium \cite{Parmar:17,article_con_bs1,app8020201}, or in a coherently driven passive microresonator \cite{bs2,bs4,PhysRevA.97.013816,PhysRevA.97.013816,prl}. \textcolor{black}{In recent papers \cite{bs3,bs4}, soliton interaction and possibility of formation of BS in microresonator system has been studied thoroughly. For normalized detuning value $\sigma<2$, the generated CS has oscillatory tail, such that a second soliton can lock at any extrema of the tail oscillations. One can find the stable and unstable equilibrium separation by calculating the interaction potential \cite{bs3,bs5}. On the other hand for higher detuning value the CS tail become monotonous, oscillations become too weak to form any BS which is the case we consider here. We encountered three different situations likely attraction, repulsion or independent soliton propagation depending on the initial delay between two CS. Our reduced analytical model can predict these results quite well.} To visualize the overall dynamics of two co-propagating CSs, we numerically solve Eq.(\ref{eq.2}) with the initial wave function of the form  $\psi(0,t)=\sqrt{2\sigma}sech(\sqrt{\sigma}(t-t_p))+ \sqrt{2\sigma}sech(\sqrt{\sigma}(t+t_p))$. We consider zero  phase modulation ($M=0$) such that $E_{in}$ becomes $P_{in}$ in Eq.(\ref{eq.2}). The simulation is performed \textcolor{black}{with the constant values of} $\sigma = 3.4$ and $P_{in} =2.3$ and in \textcolor{black}{each} case we vary the initial separation ($2t_p$) between two CSs. \subsection{\textcolor{black}{Derivation of the reduced model} }
\begin{figure}[bp]
	\centering
	\includegraphics[width=\linewidth]{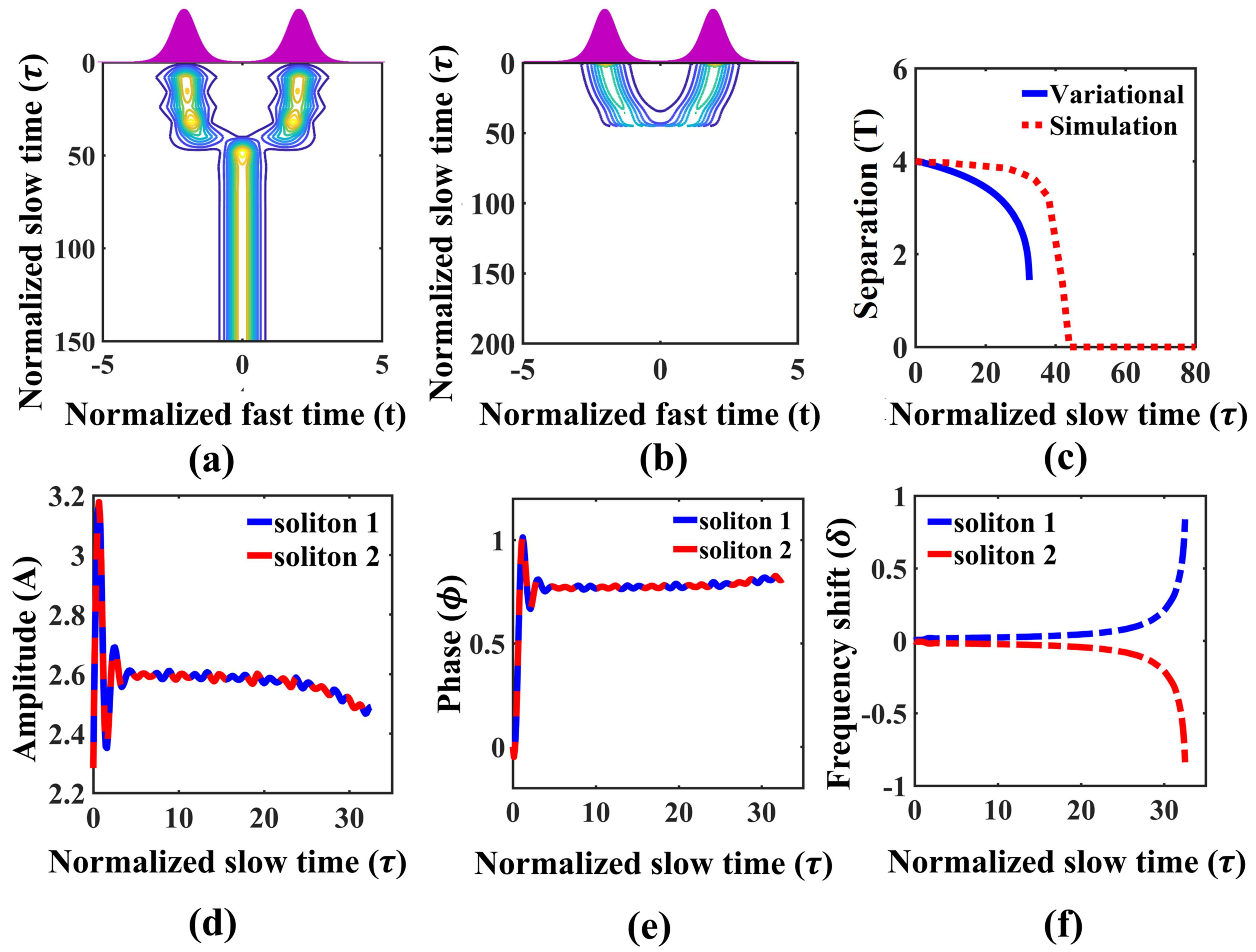}
	\caption{Evolution of two CSs when separation $2t_p = 4$. (First-row) Path of the soliton shown both (a) Numerically and (b) Variational results. (c) Separation $(2t_p = T)$ between two solitons with propagation. (Second-row) Analytical model showing evolution of two soliton parameters (d) Amplitude (e) Phase and (f) Frequency shift. }
	\label{fig.8}
\end{figure}
We find that the well-known variational treatment beautifully brings out the variation of the soliton parameters like amplitude, phase, frequency shift during different interaction scenarios. We assume soliton $1$ ($\psi_1$) propagates in such a way that only the tail of the soliton $2$ ($\psi_2$) interact with it. The governing equation of the soliton $1$ is  \textcolor{black}{\cite{vc3}}:
\begin{multline}
i\frac{\partial\psi_1}{\partial\tau}+\frac{\partial^2\psi_1}{\partial t^2} +| \psi_1|^2\psi_1 + 2|\psi_1|^2\psi_2 + \psi_1^2\psi_2^* - \sigma\psi_1 + i\psi_1\\ -i P_{in} = 0
\label{eq.18}
\end{multline}
\par The Lagrangian and the RDF correspondingly are written as:
\begin{multline}
L= \frac{i}{2}(\psi_1\psi_{1\tau}^*-\psi_{1\tau}\psi_1^*)+|\psi_{1t}|^2-\frac{1}{2}|\psi_1|^4+\sigma|\psi_1|^2
\label{eq.19}
\end{multline}

\begin{multline}
R=(2|\psi_1|^2\psi_2+\psi_1^2\psi_2^*)\psi_{1\tau}^*+ (2|\psi_1|^2\psi_2^*+\psi_1^{*2}\psi_2)\psi_{1\tau}\\+ i(\psi_1\psi_{1\tau}^*-\psi_{1\tau}\psi_1^*)- i P_{in}(\psi_{1\tau}^*-\psi_{1\tau})
\label{eq.20}
\end{multline}
\begin{figure}[bp]
	\centering
	\includegraphics[width=\linewidth]{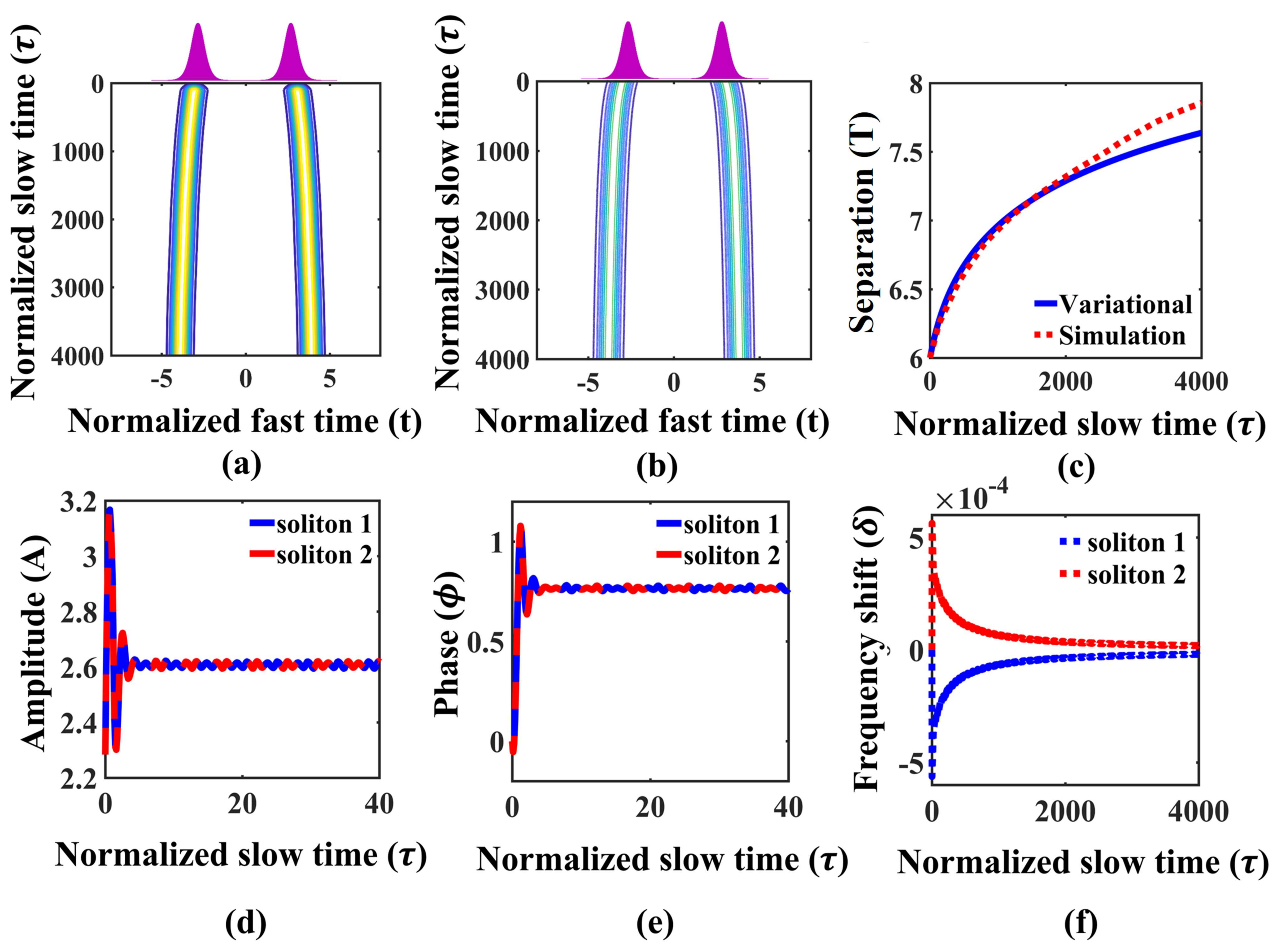}
	\caption{Evolution of two CSs when separation $2t_p = 6$. (First-row) Path of the soliton shown both (a) Numerically and (b) Variational results. (c) Separation $(2t_p = T)$ between two solitons with propagation. (Second-row) Analytical model showing evolution of two soliton parameters (d) Amplitude (e) Phase and (f) Frequency shift.}
	\label{fig.9}
\end{figure}
The derived form of the Lagrangian and RDF density using the ansatz function for soliton 1 are given in appendix C. The evolution equations for the soliton 1 \textcolor{black}{$(\psi_1)$} are:
\begin{multline}
\frac{\partial\eta_1}{\partial\tau}=-4\eta_1+16\eta_1\sqrt{\eta_1\eta_2}e^{-\sqrt{\eta_1} T}\sin\theta + \sqrt{2\eta_1}\pi P_{in}\cos\phi_1 \\-\frac{\pi^3\delta_1^2 P_{in}\cos\phi_1}{4\sqrt{2\eta_1}}
\label{eq.21}
\end{multline}
\begin{multline}
\frac{\partial\delta_1}{\partial\tau}=8\eta_1\sqrt{\eta_2}\cos\theta e^{-\sqrt{\eta_1}T}-\left(1-\frac{\pi^2\delta_1^2}{8\eta_1}\right)\times\\\frac{\delta_1 \cos\phi_1 \pi P_{in}}{\sqrt{2\eta_1}}
\label{eq.22}
\end{multline}

\begin{multline}
\frac{\partial t_1}{\partial\tau} = -2\delta_1-4\sqrt{\eta_2} \sin\theta e^{-\sqrt{\eta_1}T}+\left(1-\frac{5\pi^2\delta_1^2}{24\eta_1}\right)\times\\\frac{\pi^3 \delta_1 P_{in} \sin\phi_1} {4\sqrt{2}\eta_1^{3/2}}
\label{eq.23}
\end{multline}

\begin{multline}
\frac{\partial\phi_1}{\partial\tau} = -\sigma + \delta_1^2+ \eta_1 +12\sqrt{\eta_1\eta_2}\cos\theta e^{-\sqrt{\eta_1}T}+4\delta_1\sqrt{\eta_2}\\ \times \sin\theta e^{-\sqrt{\eta_1}T}-\frac{\pi^3 P_{in}\delta_1^2 \sin\phi_1}{2(2\eta_1)^{3/2}}\left(2-\frac{5\pi^2\delta_1^2}{24\eta_1}\right)
\label{eq.24}
\end{multline}

\textcolor{black}{where $\theta = \phi_1-\phi_2$ and $T=t_1-t_2$.}
\begin{figure}[bp]
	\centering
	\includegraphics[width=\linewidth]{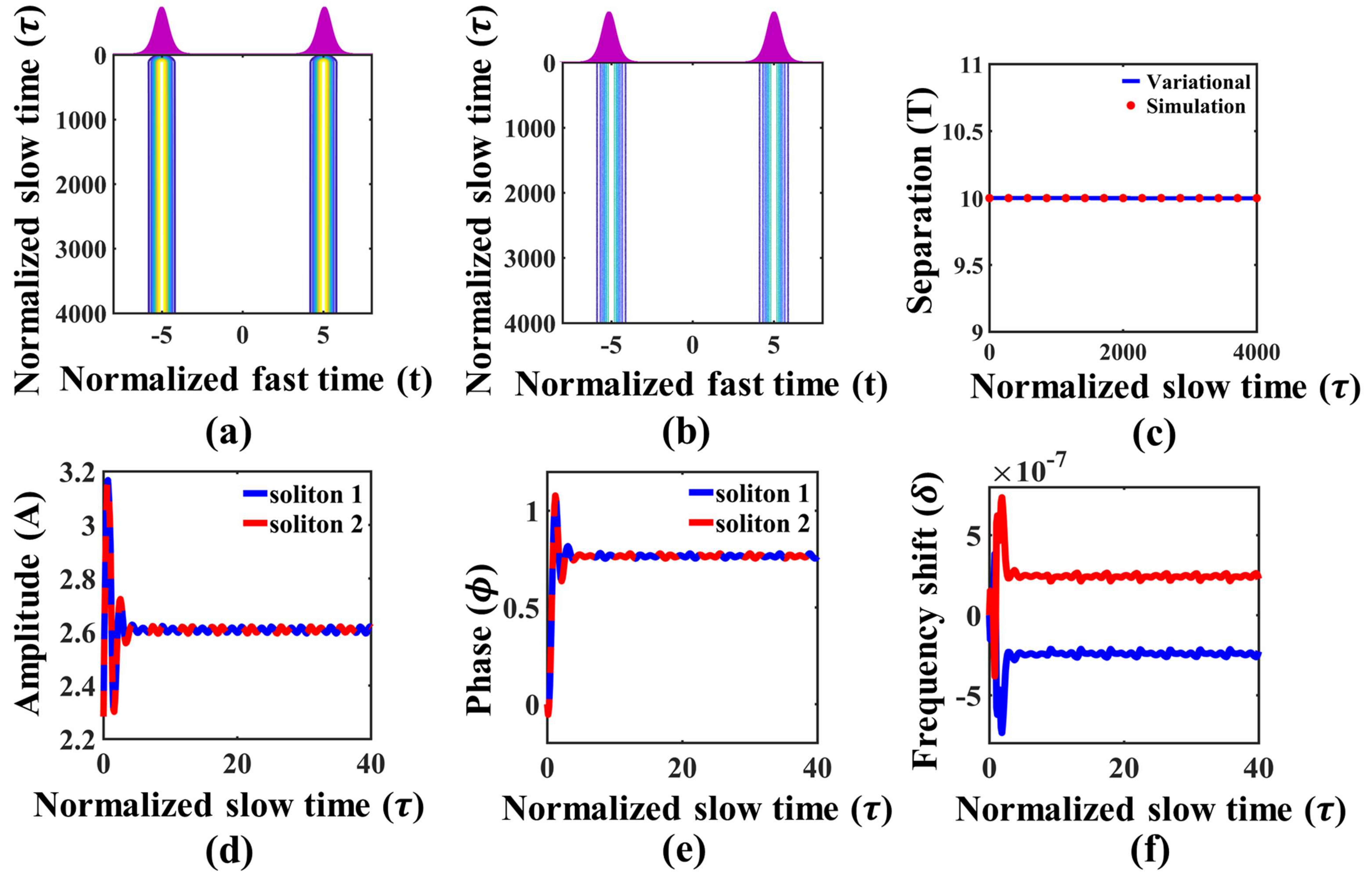}
	\caption{(First-row) Independent propagation of two CSs when the mutual separation $2t_p = 10$, (a) Numerical result and (b) Variational result. (c) Separation ($2t_p = T$) between two solitons with propagation. (Second-row) Analytical model showing evolution of two soliton parameters (d) Amplitude (e) Phase and (f) Frequency shift.}
	\label{fig.10}
\end{figure}

 \par In a similar way one can also find out the evolution equations for soliton 2 ($\psi_2$) parameters as follows:

\begin{multline}
\frac{\partial\eta_2}{\partial\tau}=-4\eta_2-16\eta_2\sqrt{\eta_1\eta_2}e^{-\sqrt{\eta_2} T}\sin\theta + \sqrt{2\eta_2}\pi P_{in}\\ \times \cos\phi_2 -\frac{\pi^3\delta_2^2 P_{in}\cos\phi_2}{4\sqrt{2\eta_2}}
\label{eq.25}
\end{multline}

\begin{multline}
\frac{\partial\delta_2}{\partial\tau}=-8\eta_2\sqrt{\eta_1}\cos\theta e^{-\sqrt{\eta_2}T}-\left(1-\frac{\pi^2\delta_2^2}{8\eta_2}\right)\times\\\frac{\delta_2 \cos\phi_2\pi P_{in}}{\sqrt{2\eta_2}}
\label{eq.26}
\end{multline}

\begin{multline}
\frac{\partial t_2}{\partial\tau} = -2\delta_2-4\sqrt{\eta_1}\sin\theta e^{-\sqrt{\eta_2}T}+\left(1-\frac{5\pi^2\delta_2^2}{24\eta_2}\right)\times\\\frac{\pi^3 \delta_2 P_{in} \sin\phi_2} {4\sqrt{2}\eta_2^{3/2}}
\label{eq.27}
\end{multline}

\begin{multline}
\frac{\partial\phi_2}{\partial\tau} = -\sigma + \delta_2^2+ \eta_2 +12\sqrt{\eta_1\eta_2}\cos\theta e^{-\sqrt{\eta_2}T}+\\4\delta_2\sqrt{\eta_1} \sin\theta e^{-\sqrt{\eta_2}T}-\frac{\pi^3 P_{in}\delta_2^2 \sin\phi_2}{2(2\eta_2)^{3/2}}\left(2-\frac{5\pi^2\delta_2^2}{24\eta_2}\right)
\label{eq.28}
\end{multline}
\begin{figure}[tp]
	\centering
	\includegraphics[width=\linewidth]{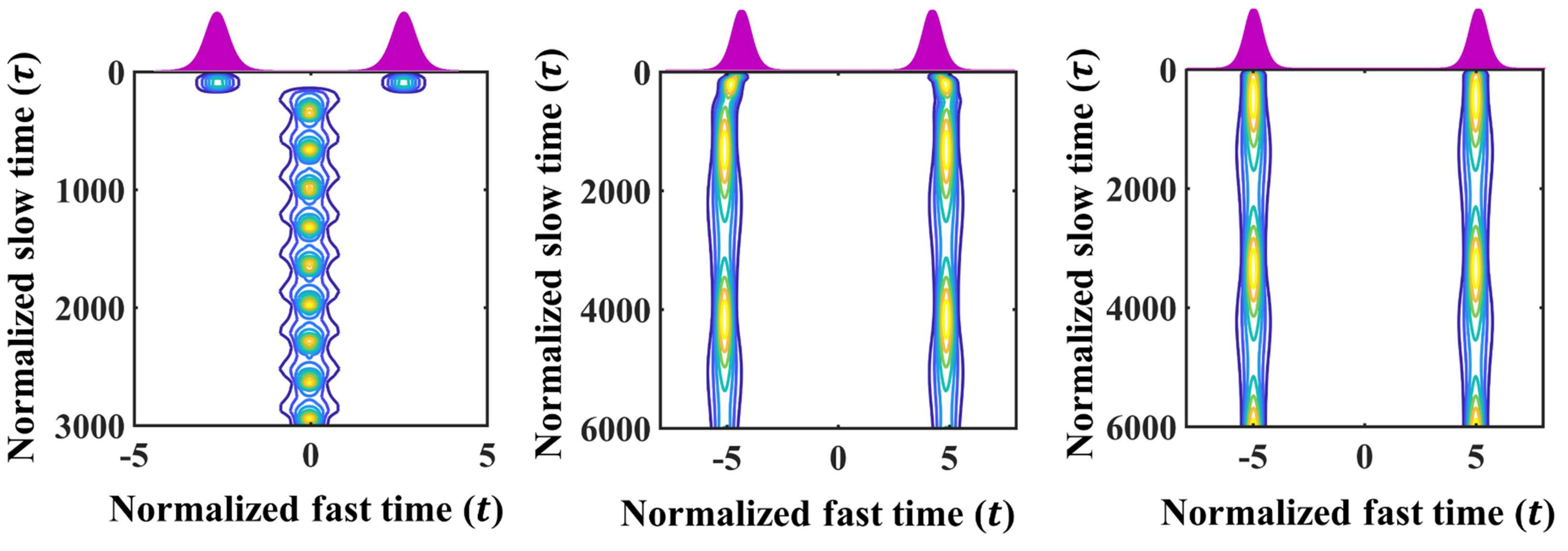}
	\caption{ Contour plot showing CS breathing state for $[\sigma,P_{in}]= [4.2,3]$ (a) Attraction state $[2t_p =4]$ (b) Repulsion state $[2t_p =6]$ (c) \textcolor{black}{Independent soliton state $[2t_p =10]$.}}
	\label{fig.11}
\end{figure}

\par A set of eight coupled ordinary differential equations (\ref{eq.21})-(\ref{eq.28}) describe the overall dynamics of two co-propagating CSs. It is observed that the initial separation between CSs controls the strength of interaction and based on which three distinct situations arise, attraction, repulsion and \textcolor{black}{independent soliton propagation.} In  Fig.(\ref{fig.8}), we show how two co-propagating CSs attract themselves to form a single state. A detailed variational treatment predicts identical behaviour (see plot \ref{fig.8}(c)) and locates the collision point from where single soliton emerges. Combining Eqs.(\ref{eq.22})-(\ref{eq.23})and Eqs.(\ref{eq.26})-(\ref{eq.27}) we derive a second-order differential equation which approximately predicts the variation of two soliton separation ($T=t_1-t_2$) given as:

\begin{equation}
\frac{\partial^{2}T}{\partial\tau^{2}}+B\frac{\partial T}{\partial\tau}\approx2A(C-2)e^{-\sqrt{\eta}T}
\label{eq.29}
\end{equation}
\begin{figure}[bp]
	\centering
	\includegraphics[width=\linewidth]{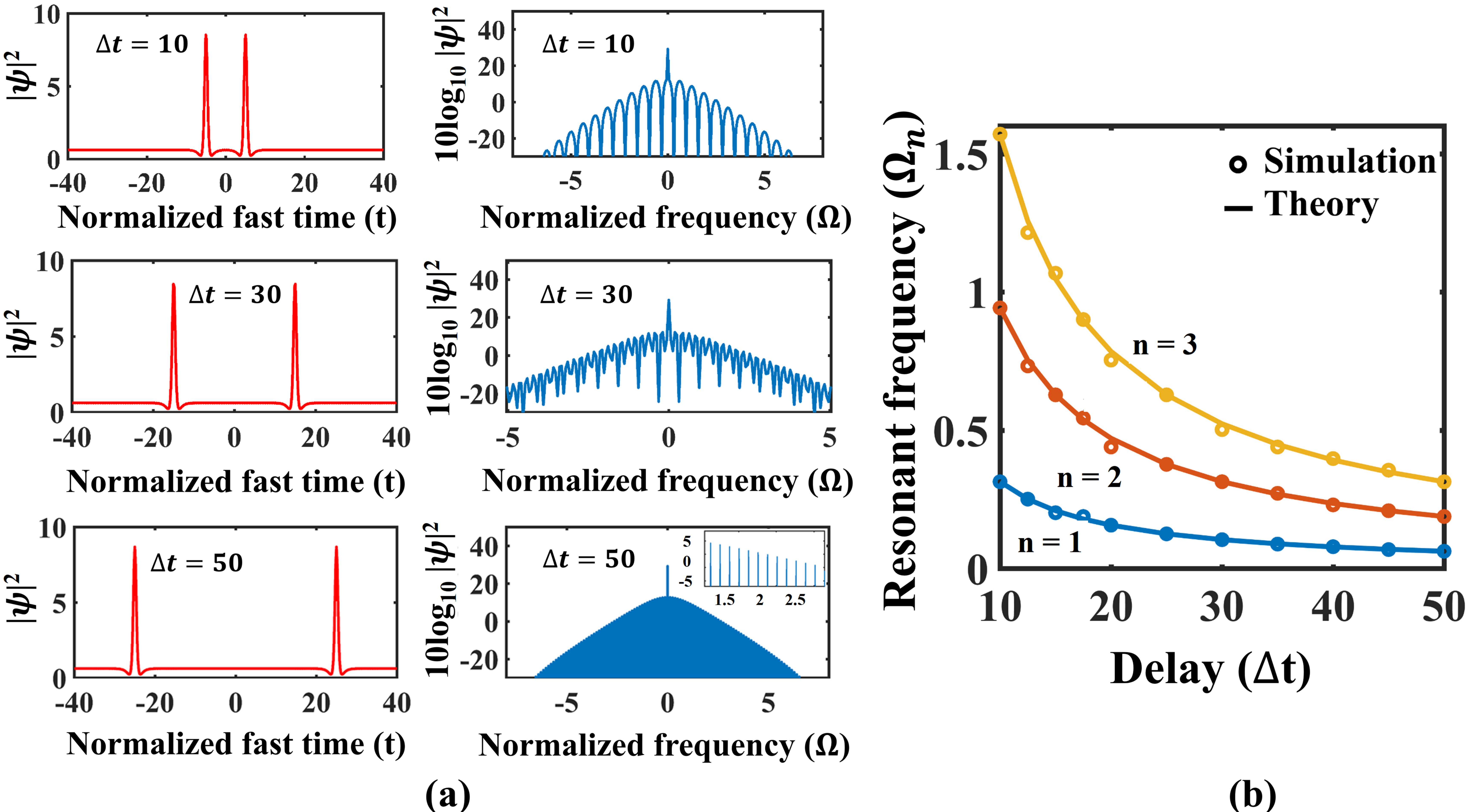}
	\caption{(a) Effect of CS interaction on generated frequency comb; interference fringe density changes with separation or delay between the two CS (b) Position of interference fringes of different order vs. delay between two CSs.}
	\label{fig.12}
\end{figure}
\begin{figure*}
	\centering
	\includegraphics[width=\linewidth]{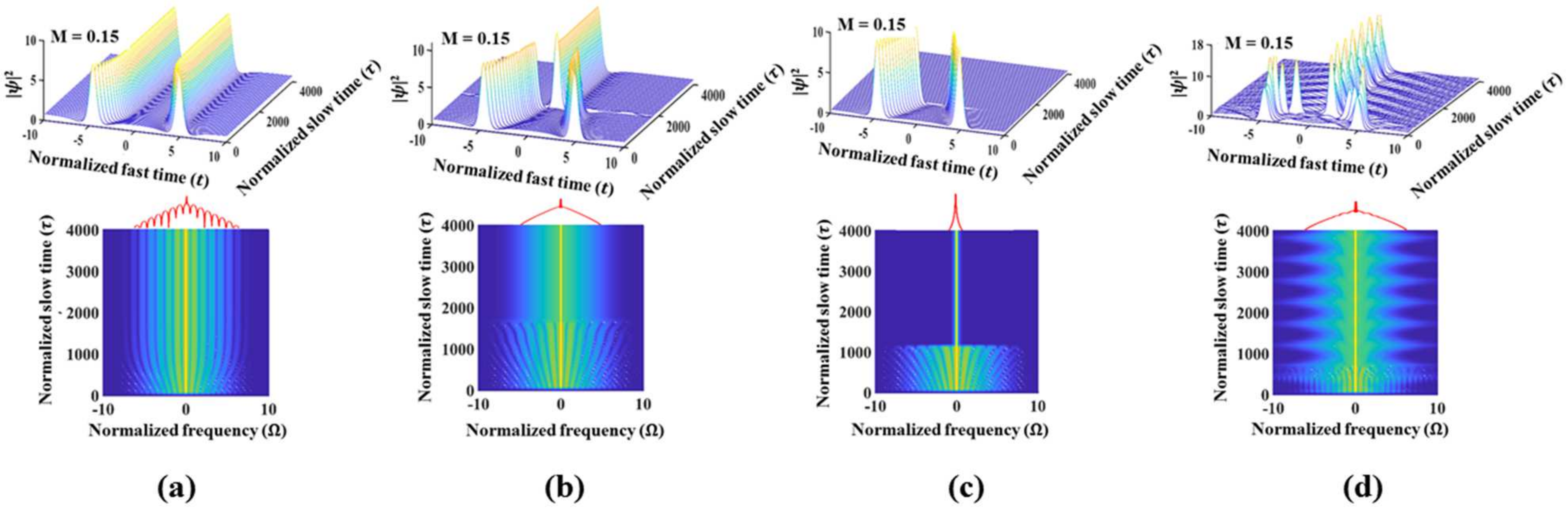}
	\caption{ Dynamics of soliton interaction in presence of cosine modulated phase profile (M = 0.15) of the driving field when the mutual separation $(2t_p) =10$ and modulation frequency $(\omega) =0.05$. (a) $[\sigma,P_{in}]= [2.9,2]$ Two soliton state, (b) $[\sigma,P_{in}]= [3.4,2.3]$ Merged CSs, (c) $[\sigma,P_{in}]= [4,2.2]$ Annihilated CSs, (d) $[\sigma,P_{in}]= [4.2,3]$ Merged breathing CSs.}
	\label{fig.13}
\end{figure*}

In order to derive this differential equation, we have assumed $\eta_1=\eta_2=\eta$ and $\phi_1=\phi_2=\phi$. The constants in the equations are as follows, $A=8\eta^{3/2}\cos\theta$, $B =\frac{\pi \cos\phi P_{in}}{\sqrt{2\eta}}$, $C = \frac{\pi^3 \sin \phi P_{in}}{(2\eta)^{3/2}}$. Note that variational analysis is applicable whenever two independent solitons exist. It will not produce any results when two solitons merge to form a single soliton because for such case, the assumption of ansatz function is violated. However, variational results work nicely before collision and we find the evolution of $A$, $\phi$ and $\delta$ in Fig. \ref{fig.8}(d-f). Further increase in separation will lead to repulsion (see Fig.\ref{fig.9})) and independent propagation of two CS (see Fig.(\ref{fig.10})). We study all these three cases based on the variational results. Note that, in all these three cases the frequency shift for individual soliton is non-identical (in fact opposite in magnitude) that results in three different dynamics. In a special situation depending upon the external parameters $(\sigma, P_{in})$, we obtain steady breather CS where the peak power of CS periodically breathes over round trip evolution. Like previous situations, different initial delay between two pulses lead to attractive or repulsive force or creation of \textcolor{black}{two independent breather CSs} (see Fig.(\ref{fig.11})).

\subsection{\textcolor{black}{Soliton interaction mediated frequency comb generation} }
It is well known that CS corresponds to frequency comb in spectral domain with single FSR spacing. Due to four wave mixing and proper phase-matching, side modes are excited from the resonant pump mode of the cavity. On the other hand for soliton interaction problem, we can intuitively say that two resonant modes are simultaneously excited which interfere.  By solving LLE, we observe interference fringes in the frequency spectrum.  Such interference fringes with high contrast reflects stable binding-separation \cite{ic2}.  Here, we choose different time delay between the two CSs where they can propagate independently. Interference patterns are changed with their binding-separation. The position of the individual interference maxima changes with the separation (see Fig.\ref{fig.12}(a)). The interference pattern becomes denser with increasing binding-separation.  For a fixed delay, the interference maximas locate at: 
\begin{equation}
\Omega_n = \pm (n-\frac{1}{2})\Delta\Omega,
\label{eq.30}
\end{equation}
where, $n= 1,2,3,..$ and $\Delta\Omega=2\pi/\Delta t $. $n$ is the interference maxima order and $\Omega_n$ are the corresponding frequencies.
In Fig.\ref{fig.12}(b) we observe that our simulation results agree quite well with the analytical formulae. 

\section{Soliton Interaction in Presence of Phase Modulated Driving Field}
\label{5}

Finally, we consider co-propagating soliton interaction  when driving beam is phase modulated with a cosine profile at a frequency $\omega$ with modulation depth $M$. To illustrate the interaction mechanism, we numerically integrate Eq.(\ref{eq.2}) using a symmetric initial condition $\psi(0,t)=\sqrt{2\eta}[sech(\sqrt{\eta}(t-t_p))+ sech(\sqrt{\eta}(t+t_p))]$. The initial separation $(2t_p=10)$ is made larger compare to the characteristic width of CS in order to avoid any self-interaction during propagation. This ensures that two CSs can move independently in absence of any modulated driving field. We observe different interaction scenarios based on two external parameters, cavity detuning frequency $(\sigma)$ and pump power $(P_{in})$. 
\begin{figure}[bp]
	\centering
	\includegraphics[width=\linewidth]{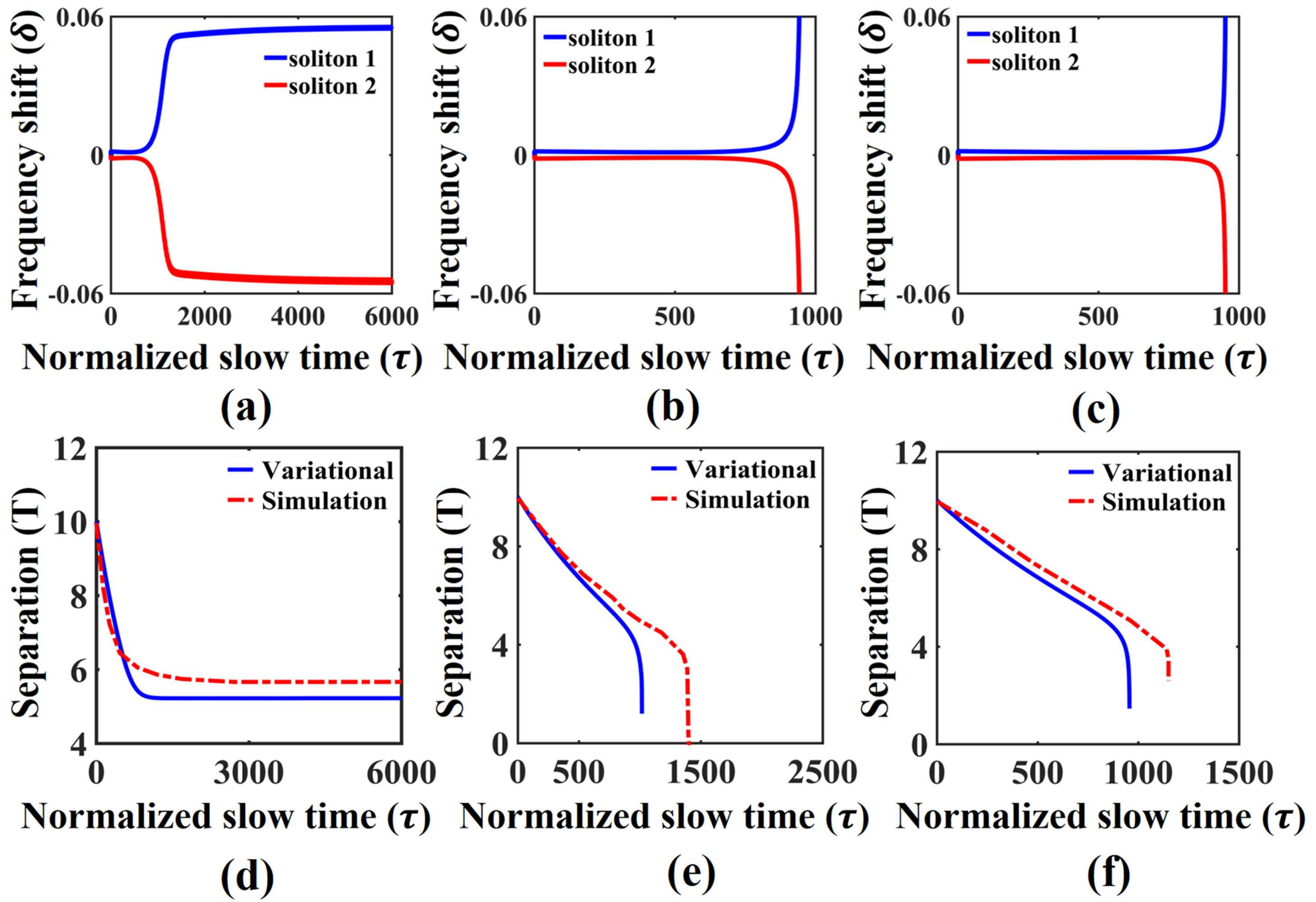}
	\caption{ (First-row) Variation of frequency-shift (variational results). (Second-row) Variation of separation of two CSs when modulation depth (M) = 0.15; (a) and (d)Two soliton state  $[\sigma,P_{in}]= [2.9,2]$, (b) and (e) Merged state $[\sigma,P_{in}]= [3.4,2.3]$, (c) and (f) Annihilated state $[\sigma,P_{in}]= [4,2.2]$.}
	\label{fig.14}
\end{figure}  
In Fig.(\ref{fig.13}), we show the evolution of two co-propagating CSs for four different sets of $(\sigma,P_{in})$. We observe that depending on the external parameters $(\sigma, P_{in})$ four distinct states may evolve, (a) a stationary two soliton state, (b) a single soliton state after merging, (c) an annihilation state and finally (d) a breathing state. The spectral evolution is also plotted for four different cases. As expected, interference fringes are observed for two soliton state whereas an annihilated state exhibits a single frequency complimenting the CW background in time domain. For single soliton state and breathing state the spectrum are almost identical.
\subsection{\textcolor{black}{Analysis of the reduced model}} 
\begin{figure}[tp]
	\centering
	\includegraphics[width=\linewidth]{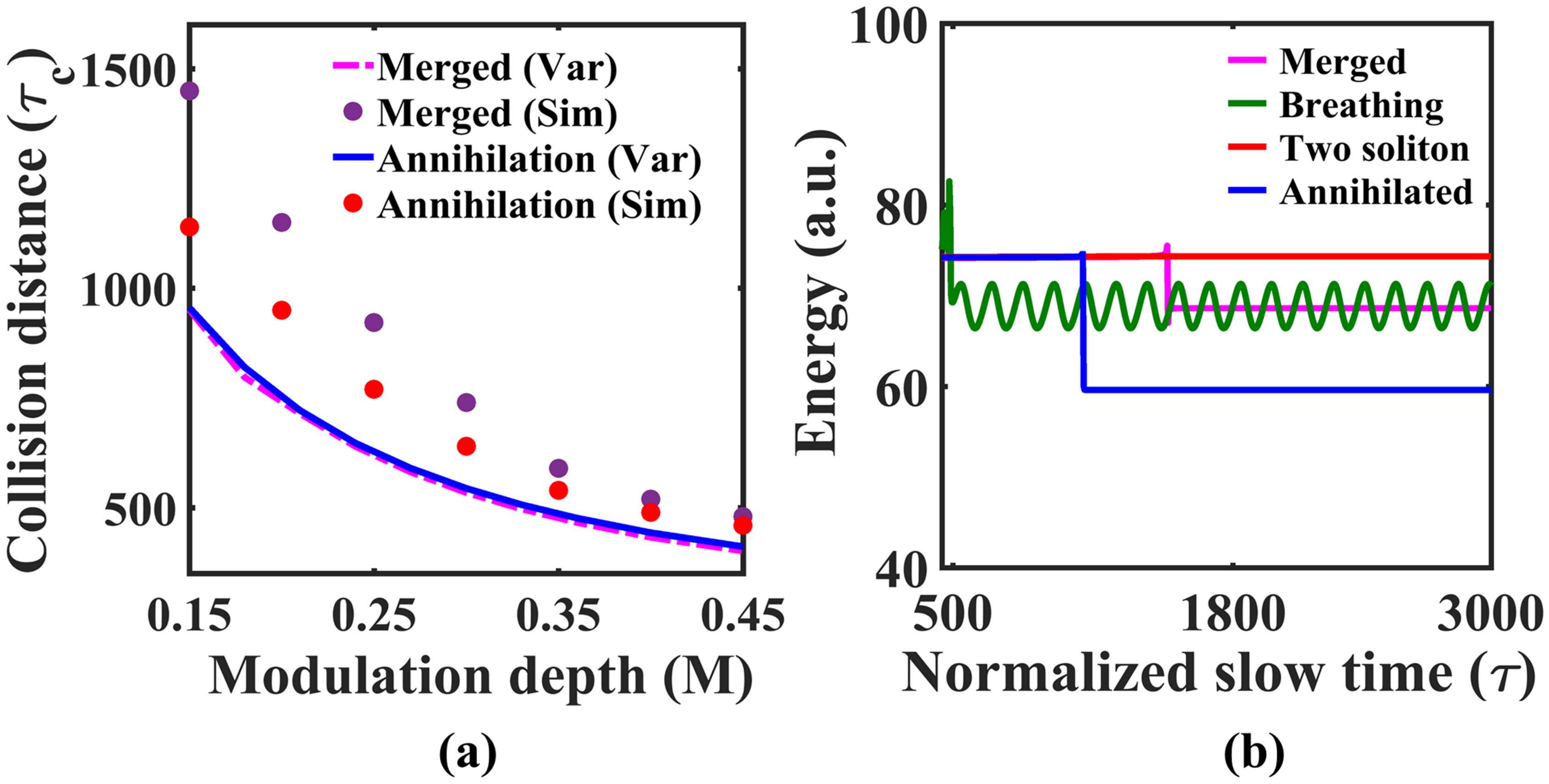}
	\caption{(a) Variation of the collision distance $(\tau_c)$ with modulation depth $(M)$, shown both with simulation and variational results for the merged and annihilated state (b) Numerically simulated intra-cavity field energy evolution for all the four (two soliton state, merged, annihilation and breathing) cases. }
	\label{fig.15}
\end{figure}

\textcolor{black}{The variational analysis discussed in earlier sections (III and IV) are useful here to visualize the pulse dynamics in this complex scenario. In Fig.(\ref{fig.14}) we observe that the different final stationary states such as two soliton state or merged single soliton state or annihilated state corresponds to different frequency shifts. Though the variational formalism can not capture the single soliton state after merging, it can well predict the point of merging and annihilation of two CSs.}  The evolution equations of the two soliton parameters are mentioned in the appendix D. We adopt the variational results (solid black lines) to obtain the separation ($T$) as a function of slow-time ($\tau$) and find a close agreement with full numerical analysis (dotted red lines). In Fig.\ref{fig.15}(a), we plot  $\tau_c$ as a function of $M$ for a fixed  $\omega$ for both merged single-soliton state and annihilated state. For a given set of $(\sigma, P_{in})$, the collision point ($\tau_c$) changes with the values of modulation depth ($M$). The solitons collide earlier for larger $M$. The simulated values are given by filled circles whereas the solid line depicts the variational result.  The variational analysis compliments the numerical results with a certain degree of accuracy. The CSs which contains a pedestal is approximated as a $sech$ shape in variational treatment and perhaps this is the reason why we have the mismatch of two results at lower modulation depth $(M)$. So, it is evident that the parameter set $(\sigma,P_{in})$ leads to different steady state of intra-cavity field where as the parameter set $(M, \omega)$ mainly controls the drift velocity of the generated CSs.
\subsection{\textcolor{black}{Intra-cavity field energy}} Finally, we compare the relative intra-cavity field energy for four cases  in Fig.\ref{fig.15}(b) by numerically computing the quantity $E=\int_{-t_{min}}^{t_{max}} \psi^*\psi\hspace{1 mm}dt$ over the slow time. We observe that total  energy remains almost steady through out the evolution time for \textcolor{black}{stationary two soliton state} (solid black line). For merged and annihilated states energy drops sharply at some critical $\tau$ corresponding to the point of merging or annihilation. During annihilation, energy does not drop to zero as the resonator always have the constant CW field, which is clearly visible in the temporal dynamics of annihilation in Fig.\ref{fig.13}(c) (the black base lines represents the CW base). For breathing state we observe oscillatory energy. The mean energy of oscillation for breathing soliton is found to be identical to the final energy of merged single soliton.
\section{Conclusion}

We have studied the complex dynamics of a single CS and co-propagating CSs separately in a passive fiber loop resonator under a phase-modulated driving field. A semi-analytical variational method is \textcolor{black}{adopted} to unfold the unique behavior of temporal CSs under this driving field. The variational calculations lead to an in-depth analysis for each perturbation situations and results in obtaining a set of ordinary differential equations which determine the evolution of individual pulse parameter. We start with a single soliton whose temporal trajectory is controlled by the modulation depth and the modulation frequency of the external pump. The variational analysis efficiently predicts the path followed by the CS. The derived explicit form of the temporal trajectory and the drift velocity of CS will enable us to control them externally with the modulation parameters. Phase-space diagram portraits the amplitude and phase evolution of CS under this perturbation. During two soliton interaction, this semi-analytical treatment can efficiently find out the role of different external parameters like delay, detuning frequency, pump power, modulation depth and modulation frequency which can enrich any numerical and experimental results. We believe that our analysis would help to investigate single or multiple CS dynamics further in the presence of several other forms of phase or amplitude modulated external field.

\begin{acknowledgments}
The author M.S would like to thank MHRD, Govt. of India and IIT Kharagpur for funding to carry out her research work.
\end{acknowledgments}

\appendix
\section{ Derivations of the reduced Lagrangian and RDF for the case mentioned in SEC. \ref{3}}
\par To obtain the set of coupled differential equations for the evolution of CS parameters, these steps are followed:
\\ \textbf{STEP 1.} \newline
The ansatz function :
\begin{multline}
\psi(t,\tau)=\sqrt{2\eta(\tau)}sech(\sqrt{\eta(\tau)}(t-t_p(\tau)))\\\times \exp[i(\phi(\tau)-\delta(\tau)(t-t_p(\tau)))],
\end{multline}
\\The driving field :
\begin{multline}
E_{in}(t) \approx P_{in}(1+i M \cos(\omega t))
\\\approx P_{in}\left(1+i M-\frac{i M \omega^2 t^2}{2}\right).  
\end{multline} The explicit form of Euler-Lagrangian (EL) eqn. is,
\begin{equation}
\frac{d}{d\tau}\left(\frac{\partial L}{\partial \psi_\tau^*}\right)+\frac{d}{d t}\left(\frac{\partial L}{\partial \psi_t^*}\right)-\frac{\partial L}{\partial \psi^*}+\left(\frac{\partial R}{\partial \psi_{\tau}^*}\right) \\= 0,
\end{equation} We choose the form of Lagrangian (L) and RDF (R) function intuitively in such a way that the above EL equation will give back the original form of LLE :$\frac{\partial \psi}{\partial \tau}=-(1+i\sigma)\psi+i|\psi|^2\psi+i\frac{\partial^{2}\psi}{\partial t^2}+E_{in}.$:
\begin{equation}
L= \frac{i}{2}(\psi\psi_\tau^*-\psi_\tau\psi^*)+|\psi_t|^2-\frac{1}{2}|\psi|^4+\sigma|\psi|^2,
\end{equation}
\begin{equation}
R=i(\psi\psi_\tau^*-\psi_\tau\psi^*)+i(E_{in}^{*}\psi_\tau-E_{in}\psi_\tau^*).
\end{equation}
\\ \textbf{STEP 2.} Now we integrate the Lagrangian and Rayleigh's function over the fast time (t):
\begin{equation}
    L_g=\int_{-\infty}^{\infty} L\hspace{1 mm}dt
\end{equation}
\begin{equation}
    R_g=\int_{-\infty}^{\infty} R\hspace{1 mm}dt
\end{equation}
\noindent The form of $(L_{g})$ and RDF $(R_{g})$ as follows, 
\begin{equation}
L_g = 4\sqrt{\eta}\left(\delta\frac{\partial t}{\partial \tau}+\frac{\partial \phi}{\partial\tau}-\frac{\eta}{3}+\sigma+\delta^2\right)
\end{equation}
\textcolor{black}{
\begin{multline}
R_g = 8\sqrt\eta\Bigl(\delta\frac{\partial t_p}{\partial \tau}+\frac{\partial \phi}{\partial \tau}\Bigr)-\frac{\partial \eta}{\partial\tau}\frac{\pi^3 P_{in}\delta^2}{\sqrt{2}\eta^2}\Bigl(\sin\phi-M \cos\phi\Bigr)\\+2\sqrt{2}P_{in} \pi\delta \frac{\partial t_p}{\partial \tau}\Bigl(1-\frac{\pi^2\delta^2}{8\eta}\Bigr)\Bigl(\cos\phi + M \sin\phi\Bigr)-\Bigl(\delta P_{in}\frac{\partial t_p}{\partial \tau}+\\
P_{in}\frac{\partial \phi}{\partial \tau}\Bigr)\Bigl(\Bigl(\cos \phi + M \sin\phi\Bigr)\Bigl(2\sqrt{2}\pi-\frac{\pi^3\delta^2}{2\sqrt{2}\eta}\Bigr)\Bigr)+\\ \frac{\delta P_{in}\pi^3}{\sqrt{2}\eta}\Bigl(1-\frac{5\pi^2\delta^2}{24\eta}\Bigr)\frac{\partial \delta}{\partial \tau}\times\Bigl(\sin\phi - M \cos\phi\Bigr)-\frac{M \omega^2}{2}\Biggl(-\frac{\partial \eta}{\partial \tau}\times \\\frac{2P_{in}\pi^3}{\eta^2}\Bigl(\cos\phi +\sqrt{2}\delta t_p \sin\phi-\frac{5\sqrt{2}\delta^2\pi^2\cos\phi}{8\eta}-\\\frac{P_{in}\delta^2 t_p^2 \cos\phi}{2\sqrt{2}}-\frac{\delta^3\pi^2 t_p \sin\phi}{6\sqrt{2}\eta}\Bigr)+\frac{\partial t_p}{\partial \tau}\Bigl(4\sqrt{2} \pi P_{in} t_p \cos\phi\Bigl(1-\\\frac{\pi^2\delta^2}{4\eta}-\frac{5\pi^4\delta^4}{96\eta^2}\Bigr) +\frac{\sqrt{2}\pi^3\delta P_{in} \sin\phi}{\eta}-\frac{5\delta^3 \sin\phi P_{in}\pi^5}{\sqrt{2}\eta^2}\Bigr)\\-\frac{\partial \phi}{\partial\tau}\Bigl(2\sqrt{2}P_{in} \sin\phi \Bigl(\frac{\pi^3}{4\eta}+\pi t_p^2\Bigr)-\sqrt{2}\delta^2 P_{in}\sin\phi \Bigl(\frac{\pi^3t_p^2}{4\eta}+\\\frac{5\pi^5}{16\eta^2}\Bigr)-\frac{\sqrt{2}P_{in}t_p\pi^3 \delta \cos\phi}{\eta}+\frac{5\sqrt{2}\pi^5\delta^3 P_{in}t_p \cos\phi }{24\eta^2}\Bigr)+\frac{\partial\delta}{\partial\tau}\\\times\Bigl(\frac{\sqrt{2}P_{in} t_p \pi^3 }{\eta} \sin\phi\Bigl(1-\frac{5\delta^2\pi^2}{4\sqrt{2}\eta}\Bigr)-2\sqrt{2}P_{in}\delta\cos\phi  \Bigl(\frac{5\pi^5}{16\eta^2}+\\\frac{\pi^3 t_p^2}{4\eta}\Bigr)+\frac{\sqrt{2}P_{in} \delta^3\cos\phi }{3\eta^2}\Bigl(\frac{61\pi^7}{64\eta}+\frac{5\pi^5 t_p^2}{16}\Bigr)\Bigr)
\end{multline}}
\\ \textbf{STEP 3.} 
 Now using the form of $L_g$ and $R_g$ in: \begin{equation}
\frac{d}{d\tau}\left(\frac{\partial L_g}{\partial \dot{p_j}}\right)-\frac{\partial L_g}{\partial p_{j}}+\left(\frac{\partial R_g}{\partial \dot{p_j}}\right) = 0,
\end{equation}
where $p_j=\eta,t_p,\phi,\delta$ and,\\\\ 
$\dot{p_{j}}= \frac{\partial\eta}{\partial\tau},\frac{\partial t_p}{\partial\tau},\frac{\partial\phi}{\partial\tau},\frac{\partial\delta}{\partial\tau}$, \textcolor{black}{we obtain the coupled ODEs for CS parameters.} These evolution equations are mentioned in Eq.(\ref{eq.9})-Eq.(\ref{eq.12}).
\vspace{3mm}
\section{\textcolor{black}{Derivation of CS trajectory under phase modulated driving field}} 
we show the simplified coupled equations including only the dominating terms required for tracing the soliton trajectory. The simplified coupled equations are:
\begin{equation}
    \frac{\partial t_p}{\partial\tau} = \Bigl(-2+ \frac{\pi^3 P_{in} \sin\phi}{4\sqrt{2}\eta^{3/2}}\Bigr)\delta-\Bigl(\frac{M\omega^2 P_{in} \pi^3 \sin\phi}{4\sqrt{2}\eta^{3/2}}\Bigr)t_p 
\end{equation}
\begin{equation}
    \frac{\partial\delta}{\partial\tau}= -\frac{\pi P_{in}\cos\phi}{\sqrt{2\eta}}\delta+\frac{M\omega^2\pi P_{in}\cos\phi}{\sqrt{2\eta}}t_p
\end{equation}\\ We can write it the form:
\begin{equation}
    \frac{\partial t_p}{\partial\tau} = A_{1}\delta-A_{2}t_p
\end{equation}
\begin{equation}
    \frac{\partial\delta}{\partial\tau} = -B_{1}\delta+B_{2}t_p
\end{equation}\\ where, $A_{1} = \Bigl(-2+ \frac{\pi^3 P_{in} \sin\phi}{4\sqrt{2}\eta^{3/2}}\Bigr), A_{2}= \Bigl(\frac{M\omega^2 P_{in} \pi^3 \sin\phi}{4\sqrt{2}\eta^{3/2}}\Bigr), B_{1} = \frac{\pi P_{in}\cos\phi}{\sqrt{2\eta}}, B_{2}= \frac{M\omega^2\pi P_{in}\cos\phi}{\sqrt{2\eta}}$.
The set of coupled differential equation can be easily decoupled and we can get a second order differential equation of $t_p$ of the form:
\begin{equation}
    \frac{\partial^2 t_p}{\partial\tau^2}+D_{1}\frac{\partial t_p}{\partial\tau}+D_{2}t_p = 0
\end{equation}\\ where, $D_{1} = (A_2+B_1)=\frac{M\omega^2 P_{in}\pi^3 \sin\phi}{4\sqrt{2}\eta^{3/2}}+\frac{\pi P_{in} \cos\phi }{\sqrt{2\eta}}, D_{2}=(B_1A_2-A_1B_2) = \frac{\sqrt{2}M\omega^2\pi P_{in} \cos\phi}{\sqrt{\eta}}$. Now, the solution of the above equation can  simply be written as, $t_p(\tau)=Xe^{m_1\tau}+Ye^{m_2\tau}$, where $m_{1,2}=[-D_1\pm D_1(1-4\Delta)^{1/2}]/2$, with $\Delta=D_2/D_1^2$ and $X, Y$ are constants. Note that, $\Delta$ is very small (as $M\omega^2 <<1$) then $m_1 \approx -D_2/D_1$ and $m_2 \approx -D_1(1-\Delta)$. 
Now the boundary conditions $t_p(0)=t_0$ and $v_d(0)=0$ leads to $X=t_0(1+\Delta)$ and $Y=-\Delta t_0$. Now neglecting  $\Delta$ and $A_2$ we may have the following approximate expression of the temporal position,
\begin{equation}
    t_p \approx t_{0}\exp(-2M\omega^2\tau).
\end{equation}
\vspace{3mm}
\section{Calculations for the reduced Lagrangian and RDF in case mentioned in SEC. \ref{4}} 
To obtain the evolution equations of two CSs we follow the same procedure mentioned in \textbf{Appendix A}. \\ \textbf{STEP 1:} 
 For two-soliton interaction problem we choose our ansatz as follows: 
 \\For soliton $1$ :
\begin{multline}
\psi_1(t,\tau)=\sqrt{2\eta_1(\tau)}sech(\sqrt{\eta_1(\tau)}(t-t_1(\tau)))\times\\ \exp\Biggl(i[\phi_1(\tau)-\delta_1(\tau)(t-t_1(\tau))]\Biggr)
\end{multline}
The interaction between two solitons is mediated with the tail oscillations, therefore we approximate $\psi_2$ in the following form:
\begin{multline}
\psi_2(t,\tau) \approx 2\sqrt{2\eta_2}\exp\Biggl(-\sqrt{\eta_2}(t-t_2(\tau))\Biggr)\times\\\exp\Biggl(i[\phi_2(\tau)-\delta_2(\tau)(t-t_2(\tau))]\Biggr)
\end{multline} The explicit form of Euler-Lagrangian (EL) eqn. is
\begin{equation}
\frac{d}{d\tau}\left(\frac{\partial L}{\partial \psi_{1\tau}^*}\right)+\frac{d}{d t}\left(\frac{\partial L}{\partial \psi_{1t}^*}\right)-\frac{\partial L}{\partial \psi_1^*}+\left(\frac{\partial R}{\partial \psi_{1\tau}^*}\right) \\= 0,
\end{equation}
Considering $\psi = \psi_1+\psi_2$, we can write the LLE for $\psi_1$ as :
\begin{multline}
i\frac{\partial\psi_1}{\partial\tau}+\frac{\partial^2\psi_1}{\partial t^2} +| \psi_1|^2\psi_1 + 2|\psi_1|^2\psi_2 + \psi_1^2\psi_2^* - \sigma\psi_1 + i\psi_1 \\-i P_{in} = 0
\end{multline} We choose the form of Lagrangian (L) and RDF (R) function in such a way that the EL equation for $\psi_1$ will give back the the LLE for $\psi_1$ as mentioned in C4:
\begin{multline}
L= \frac{i}{2}(\psi_1\psi_{1\tau}^*-\psi_{1\tau}\psi_1^*)+|\psi_{1t}|^2-\frac{1}{2}|\psi_1|^4+\sigma|\psi_1|^2
\end{multline}
\begin{multline}
R=(2|\psi_1|^2\psi_2+\psi_1^2\psi_2^*)\psi_{1\tau}^*+ (2|\psi_1|^2\psi_2^*+\psi_1^{*2}\psi_2)\psi_{1\tau}\\+ i(\psi_1\psi_{1\tau}^*-\psi_{1\tau}\psi_1^*)- i P_{in}(\psi_{1\tau}^*-\psi_{1\tau})
\end{multline}
\textbf{STEP 2.} Now we integrate the Lagrangian and Rayleigh's function over the fast time (t):
\begin{equation}
    L_g=\int_{-\infty}^{\infty} L\hspace{1 mm}dt
\end{equation}
\begin{equation}
    R_g=\int_{-\infty}^{\infty} R\hspace{1 mm}dt
\end{equation}
\\\noindent The form of $(L_{g})$ and RDF $(R_{g})$ as follows, 
\begin{equation}
L_g = 4\sqrt{\eta_1}\left(\delta_1\frac{\partial t_1}{\partial\tau}+\frac{\partial\phi_1}{\partial\tau_1}-\frac{\eta_1}{3}+\sigma +\delta_1^2 \right)
\end{equation}
\begin{multline}
R_g = 8\sqrt{\eta_1}\Bigl(\delta_1\frac{\partial t_1}{\partial\tau}+ \frac{\partial \phi_1}{\partial\tau}\Bigr)+ 8\eta_1^{3/2}\eta_2^{1/2}e^{-\sqrt{\eta_2}T}\Biggl(6\cos\theta\times\\\Bigl(\frac{1}{2\eta_1^{3/2}}\frac{\partial\eta_1}{\partial\tau}-\frac{2}{3}\frac{\partial t_1}{\partial \tau}\Bigr)-2\sin\theta\times\Bigl(\frac{2\delta_1}{\sqrt{\eta_1}}\frac{\partial t_1}{\partial\tau}+\frac{2}{\sqrt{\eta_1}}\frac{\partial \phi_1}{\partial\tau}+\\\frac{1}{\eta_1}\frac{\partial\delta_1}{\partial\tau}\Bigr)\Biggr)-\frac{\pi^3P_{in}\delta_1^2\sin\phi_1}{2\sqrt{2}\eta_1^2}\frac{\partial\eta_1}{\partial\tau}+\Bigl(1-\frac{\pi^2\delta_1^2}{8\eta_1}\Bigr)2\sqrt{2} P_{in}\pi\delta_1\\\times \cos\phi_1\frac{\partial t_1}{\partial\delta_1}+\frac{\pi^3\delta_1P_{in}\sin\phi_1}{\sqrt{2}\eta_1}\Bigl(1-\frac{5\pi^2\delta_1^2}{24\eta_1}\Bigr)-\Bigl(\delta_1 P_{in}\frac{\partial t_1}{\partial \tau}\\+ P_{in}\frac{\partial\phi_1}{\partial\tau}\Bigr)\Bigl(2\sqrt{2}\pi \cos\phi_1 -\frac{\pi^3\delta_1^2 \cos\phi_1}{2\sqrt{2}\eta_1}\Biggr)
\end{multline}
\noindent where $\theta = \phi_1-\phi_2$ and $T=t_1-t_2$. 
\\ \textbf{STEP 3.} 
 Now using the form of $L_g$ and $R_g$ in: \begin{equation}
\frac{d}{d\tau}\left(\frac{\partial L_g}{\partial \dot{p_j}}\right)-\frac{\partial L_g}{\partial p_{j}}+\left(\frac{\partial R_g}{\partial \dot{p_j}}\right) = 0,
\end{equation}
where $p_j=\eta_1,t_1,\phi_1,\delta_1$ and,\\\\ 
$\dot{p_{j}}= \frac{\partial\eta_1}{\partial\tau},\frac{\partial t_1}{\partial\tau},\frac{\partial\phi_1}{\partial\tau},\frac{\partial\delta_1}{\partial\tau}$, we obtain the coupled ODEs for CS parameters. These evolution equations are mentioned in Eq.(\ref{eq.21})-Eq.(\ref{eq.24}).\\\\ \par \textcolor{black}{Similarly to obtain the evolution equation for the second CS, which was mentioned in Eq.(\ref{eq.25})-Eq.(\ref{eq.28}), we have interchanged the form of $\psi_1$ and $\psi_2$ in the ansatz function mentioned in (C1) and (C2) and follow the same procedure to obtain the coupled ODEs for the parameters of second CS.} 

\section{\textcolor{black}{Coupled differential equations of two soliton parameters mentioned in SEC. \ref{5}}} 
Following the procedures mentioned in \textbf{Appendix A} for single CS propagation under phase-modulated driving field and in \textbf{Appendix C} two soliton propagation with constant driving field one can easily obtain the governing equations of two CS parameters under phase modulated driving field. Here are the coupled mode equations for the two CSs:
\begin{multline}
\frac{\partial\eta_1}{\partial\tau}=-4\eta_1+16\eta_1\sqrt{\eta_1\eta_2}\exp(-\sqrt{\eta_1}T)\sin\theta+ \frac{\sqrt{\eta_1}}{2}\times\\\Bigl(2\sqrt{2}\pi P_{in}-\frac{\pi^3\delta_1^2 P_{in}}{2\sqrt{2}\eta_1}\Bigr)\Bigl(\cos\phi_1+M\sin\phi_1\Bigr)-\frac{M\omega^2\sqrt{\eta_1}}{4}\times\\\Biggr(2\sqrt{2}P_{in}\sin\phi_1\left(\frac{\pi^3}{4\eta_1}+\pi t^2_1\right)-\frac{2\delta_1^2 P_{in} }{\eta_1}\sin\phi_1\Bigl(\frac{\pi^3 t_1^2}{4}+\\\frac{5 \pi^5}{16\eta_1}\Bigr)-\frac{\sqrt{2}\delta_1 \pi^3 P_{in} t_1 \cos\phi_1}{\eta_1}+\frac{5\sqrt{2}\pi^5\delta_1^3 P_{in} }{24\eta_1^2}t_1 \cos\phi_1\Biggr)
\end{multline}
\begin{multline}
\frac{\partial\phi_1}{\partial\tau}=-\sigma+\eta_1+\delta_1^2+12\sqrt{\eta_1\eta_2}\cos\theta \exp(-\sqrt{\eta_2}T)+\\4\delta_1\sqrt{\eta_2}\sin\theta \exp(-\sqrt{\eta_1}T)+\frac{\pi^3P_{in} \delta_1^2}{2\sqrt{2}\eta_1^{3/2}}\Biggl(\sin\phi_1 +M \cos\phi_1\Biggr) \\+\frac{5\sqrt{2}M\omega^2\pi^3\delta_1 P_{in}t_1 \sin\phi_1}{8\eta_1^{3/2}}-\frac{5\pi^5\delta_1^4 P_{in} M \cos\phi_1}{96\sqrt{2}\eta_1^{3/2}}-\\\frac{55\sqrt{2}\delta_1^2 P_{in} t_1 \pi^5 M\omega^2 \sin\phi_1}{192\eta_1^{3/2}}-\frac{M\omega^2 P_{in}\delta_1^4 \cos\phi_1}{12\sqrt{2}\eta_1^{3/2}}\Biggl(\frac{61\pi^7}{64\eta_1}\\+\frac{5\pi^5 t_1^2}{16}\Biggr)-\frac{M \omega^2 P_{in} \delta_1^2 \cos\phi_1 \pi^3}{8\eta_1^{3/2}}\Biggl(\frac{25\sqrt{2}\pi^2}{8\eta_1}+\frac{3t_1^2}{\sqrt{2}}\Biggr)
\end{multline}
\begin{multline}
\frac{\partial\delta_1}{\partial\tau}=8\eta_1\sqrt{\eta_2}\cos\theta \exp(-\sqrt{\eta_1}T)-\frac{\delta_1}{4\sqrt{\eta_1}}\Bigl(2\sqrt{2}\pi P_{in}\\-\frac{\pi^3\delta_1^2 P_{in}}{2\sqrt{2}\eta_1}\Bigr)\Bigl(\cos\phi_1+M \sin\phi_1\Bigr)+ \frac{M\omega^2}{4\sqrt{\eta_1}}\Biggl(2\sqrt{2}\pi P_{in}t_1  \cos\phi_1\\+\sqrt{2}\delta_1 P_{in}\pi t_1^2 \sin\phi_1 \Bigl(1- \frac{\delta_1^2\pi^2}{4\sqrt{2}\eta_1}\Bigr)+\frac{3\delta_1 P_{in}\pi^3 \sin\phi_1}{2\sqrt{2}\eta_1}\\-\Bigl(1+\frac{5\pi^2}{96\eta_1}\Bigr)\frac{\sqrt{2}\pi^3\delta_1^2 P_{in}t_1 \cos\phi_1}{\eta_1}\Biggr)
\end{multline}
\begin{multline}
\frac{\partial t_1}{\partial\tau}=-2\delta_1-4\sqrt{\eta_2}\sin\theta \exp (-\sqrt{\eta_1}T)+\frac{\delta_1 P_{in}\pi^3}{4\sqrt{2}\eta_1^{3/2}}\Bigl(\sin\phi_1\\-M \cos\phi_1\Bigr)-\frac{M\omega^2}{4\sqrt{2\eta_1^{3}}} \Biggl(P_{in}t_1\pi^3 \sin\phi_1-2\sqrt{2}\delta_1 P_{in}\cos\phi_1\\\times \Biggl(\frac{5\pi^5}{16\eta_1}+\frac{\pi^3t_1^2}{4}\Biggr)-\frac{5\pi^5\delta_1^2 P_{in}t_1 \sin\phi_1}{8\sqrt{2}\eta_1}+\frac{P_{in} \delta_1^3 \cos\phi_1}{3\sqrt{2}\eta_1}\\\times\Biggl(\frac{61\pi^7}{64\eta_1}+\frac{5\pi^5t_1^2}{16}\Biggr)\Biggr)
\end{multline}
\noindent Here, $\eta_1(\eta_2), t_1(t_2), \phi_1(\phi_2), \delta_1(\delta_2)$ are the corresponding amplitude, position, phase and frequency shift of the soliton 1( soliton 2), $\theta = \phi_1-\phi_2$ and $T=t_1-t_2$. 
\par The evolution equations for the second CS are :
\begin{multline}
\frac{\partial\eta_2}{\partial\tau}=-4\eta_2-16\eta_2\sqrt{\eta_2\eta_1}\exp(-\sqrt{\eta_2}T)\sin\theta+ \frac{\sqrt{\eta_2}}{2}\times\\\Bigl(2\sqrt{2}\pi P_{in}-\frac{\pi^3\delta_2^2 P_{in}}{2\sqrt{2}\eta_2}\Bigr)\Bigl(\cos\phi_2+M\sin\phi_2\Bigr)-\frac{M\omega^2\sqrt{\eta_2}}{4}\times\\\Biggr(2\sqrt{2}P_{in}\sin\phi_2\left(\frac{\pi^3}{4\eta_2}+\pi t^2_2\right)-\frac{2\delta_2^2 P_{in} }{\eta_2}\sin\phi_2\Bigl(\frac{\pi^3 t_2^2}{4}+\\\frac{5 \pi^5}{16\eta_2}\Bigr)-\frac{\sqrt{2}\delta_2 \pi^3 P_{in} t_2 \cos\phi_2}{\eta_2}+\frac{5\sqrt{2}\pi^5\delta_2^3 P_{in} }{24\eta_2^2}t_2 \cos\phi_2\Biggr)
\end{multline}
\begin{multline}
\frac{\partial\phi_2}{\partial\tau}=-\sigma+\eta_2+\delta_2^2+12\sqrt{\eta_2\eta_1}\cos\theta \exp(-\sqrt{\eta_1}T)+\\4\delta_2\sqrt{\eta_1}\sin\theta \exp(-\sqrt{\eta_2}T)+\frac{\pi^3P_{in} \delta_2^2}{2\sqrt{2}\eta_2^{3/2}}\Biggl(\sin\phi_2 +M \cos\phi_2\Biggr) \\+\frac{5\sqrt{2}M\omega^2\pi^3\delta_2 P_{in}t_2 \sin\phi_2}{8\eta_2^{3/2}}-\frac{5\pi^5\delta_2^4 P_{in} M \cos\phi_2}{96\sqrt{2}\eta_2^{3/2}}-\\\frac{55\sqrt{2}\delta_2^2 P_{in} t_2 \pi^5 M\omega^2 \sin\phi_2}{192\eta_2^{3/2}}-\frac{M\omega^2 P_{in}\delta_2^4 \cos\phi_2}{12\sqrt{2}\eta_2^{3/2}}\Biggl(\frac{61\pi^7}{64\eta_2}\\+\frac{5\pi^5 t_2^2}{16}\Biggr)-\frac{M \omega^2 P_{in} \delta_2^2 \cos\phi_2 \pi^3}{8\eta_2^{3/2}}\Biggl(\frac{25\sqrt{2}\pi^2}{8\eta_2}+\frac{3t_2^2}{\sqrt{2}}\Biggr)
\end{multline}
\begin{multline}
\frac{\partial\delta_2}{\partial\tau}=-8\eta_2\sqrt{\eta_1}\cos\theta \exp(-\sqrt{\eta_2}T)-\frac{\delta_2}{4\sqrt{\eta_2}}\Bigl(2\sqrt{2}\pi P_{in}\\-\frac{\pi^3\delta_2^2 P_{in}}{2\sqrt{2}\eta_2}\Bigr)\Bigl(\cos\phi_2+M \sin\phi_2\Bigr)+ \frac{M\omega^2}{4\sqrt{\eta_2}}\Biggl(2\sqrt{2}\pi P_{in}t_2  \cos\phi_2\\+\sqrt{2}\delta_2 P_{in}\pi t_2^2 \sin\phi_2 \Bigl(1- \frac{\delta_2^2\pi^2}{4\sqrt{2}\eta_2}\Bigr)+\frac{3\delta_2 P_{in}\pi^3 \sin\phi_2}{2\sqrt{2}\eta_2}\\-\Bigl(1+\frac{5\pi^2}{96\eta_2}\Bigr)\frac{\sqrt{2}\pi^3\delta_2^2 P_{in}t_2 \cos\phi_2}{\eta_2}\Biggr)
\end{multline}
\begin{multline}
\frac{\partial t_2}{\partial\tau}=-2\delta_2-4\sqrt{\eta_1}\sin\theta \exp (-\sqrt{\eta_2}T)+\frac{\delta_2 P_{in}\pi^3}{4\sqrt{2}\eta_2^{3/2}}\Bigl(\sin\phi_2\\-M \cos\phi_2\Bigr)-\frac{M\omega^2}{4\sqrt{2\eta_2^{3}}} \Biggl(P_{in}t_2\pi^3 \sin\phi_2-2\sqrt{2}\delta_2 P_{in}\cos\phi_2\\\times \Biggl(\frac{5\pi^5}{16\eta_2}+\frac{\pi^3t_2^2}{4}\Biggr)-\frac{5\pi^5\delta_2^2 P_{in}t_2 \sin\phi_2}{8\sqrt{2}\eta_2}+\frac{P_{in} \delta_2^3 \cos\phi_2}{3\sqrt{2}\eta_2}\\\times\Biggl(\frac{61\pi^7}{64\eta_2}+\frac{5\pi^5t_2^2}{16}\Biggr)\Biggr)
\end{multline}
These set of eight coupled differential equations (D1-D8) help to visualize the evolution of two CS parameters under phase-modulated driving field.
\noindent


\begin{thebibliography}{99}
\bibitem{cs1}
T. Herr, V. Brasch, J.D. Jost, C.Y. Wang, N.M. Kondratiev, M. L. Gorodetsky, and T. J. Kippenberg, Nat. Photonics \textbf{8}, 145  (2014).

\bibitem{cs2}
J. K. Jang, M. Erkintalo, S. Coen, and S. G. Murdoch, Nat. Commun. \textbf{6:7370} (2015).

\bibitem{cs3}
P. Grelu, \textit{Nonlinear Optical Cavity Dynamics: From Microresonators to Fiber Lasers},
1st ed.
(Wiley-VCH,
2016).
\bibitem{inbook}

Y. Kivshar and G. P.  Agrawal , \textit{Optical Solitons: From Fibers to Photonic Crystals}, (Academic Press,
2003).

\bibitem{in1book}
A. Ankiewicz and N. Akhmediev, \textit{Dissipative Solitons: From Optics to Biology and Medicine}, (Springer, Berlin, Heidelberg, 2008).

\bibitem{fo}

F. Leo, S. Coen, P. Kockaert, S.-P. Gorza and P. Emplit, and M. Haelterman, Nat. Photonics
\textbf{4}, 471 (2010).

\bibitem{of1}
A. Roy, R. Haldar, and S. K. Varshney,
J. Lightwave Tech. \textbf{36}, 5807 (2018).

\bibitem{of2}
J. K. Jang, M. Erkintalo, J. Schr\"{o}der, B. J. Eggleton, S. G. Murdoch, and S. Coen, Opt. Lett. \textbf{41}, 4526 (2016).


\bibitem{fc1}
A. G. Griffith, R. K. Lau, J. Cardenas, Y. Okawachi, A. Mohanty, R. Fain, Y. H. Daniel Lee, M. Yu, C. T. Phare, C. B. Poitras, A. L. Gaeta, and M. Lipson, Nat. Commun.
\textbf{6:6299} (2015).
\bibitem{fc2}
S. Miller, K. Luke, Y. Okawachi, J. Cardenas, A. L. Gaeta, and M. Lipson, Opt. Express
\textbf{22} 26517 (2014).

\bibitem{pm1}

H. Taheri, A. A. Eftekhar, K. Wiesenfeld, and A. Adibi, IEEE Photon. J. \textbf{7}, 1 (2015).

\bibitem{pm2}

J. K. Jang, M. Erkintalo, S. G. Murdoch, and S. Coen, Opt. Lett. \textbf{40}, 4755 (2015). 
\bibitem{pm3}
W. J. Firth and A. J. Scroggie, Phys. Rev. Lett. \textbf{76}, 1623 (1996).

\bibitem{doi:10.1063/1.2828458}
F. Pedaci, S. Barland, E. Caboche, P. Genevet, M. Giudici, J. R. Tredicce, T. Ackemann, A. J. Scroggie, W. J. Firth, G.-L. Oppo, G. Tissoni, and R. Jäger, App. Phys. Lett. \textbf{92}, 011101 (2008).

\bibitem{doi:10.1063/1.2388867}

F. Pedaci, P. Genevet, S. Barland, M. Giudici, and J. R. Tredicce, App. Phys. Lett. \textbf{89}, 221111 (2006).

\bibitem{vc1}
D. Anderson, Phys. Rev. A \textbf{27}, 3135 (1983).
\bibitem{vc2}
S. Roy, S. K. Bhadra, and G. P. Agrawal, Opt. Comm. \textbf{281}, 5889 (2008).

\bibitem{vc3}

G. P. Agrawal, \textit{Nonlinear Fiber Optics}, 5th ed. (Academic Press, 2012).

\bibitem{var_dis2}
S. C. Cerda, S.B. Cavalcanti, 
and J.M. Hickmann, Eur. Phys. J. D \textbf{1}, 313 (1998).

\bibitem{vd1}

A. Sahoo, S. Roy, and G. P. Agrawal, Phys. Rev. A \textbf{96}, 013838 (2017).

\bibitem{vd2}

N. N. Akhmediev, A. Ankiewicz, and J. M. Soto-Crespo, Phys. Rev. Lett. \textbf{79}, 4047 (1997).

\bibitem{bs2}
S. Wabnitz, Opt. Lett. \textbf{18}, 601 (1993).

\bibitem{var_dis1}
X. Yi, Q. F. Yang, K. Y. Yang and K. Vahala, Opt. Lett. \textbf{41}, 3419 (2016).

\bibitem{var-dissipative}
W. B. Cardoso, L. Salasnich, and B. A. Malomed, Sci. Rep. \textbf{7}, 850 (2017).

\bibitem{l1}
L. A. Lugiato and R. Lefever, Phys. Rev. Lett. \textbf{58}, 2209 (1987).

\bibitem{l2}
I. Hendry, W. Chen, Y. Wang, B. Garbin, J.  Javaloyes, G.-L. Oppo, S. Coen, S. G. Murdoch, and M. Erkintalo, Phys. Rev. A \textbf{97}, 053834 (2018).

\bibitem{PhysRevA.82.033801}
Y. K. Chembo and N. Yu, Phys. Rev. A \textbf{82}, 033801 (2010).

\bibitem{PhysRevA.89.063814}
C. Godey, I. V. Balakireva, A. Coillet, and Y. K. Chembo, Phys. Rev. A \textbf{89}, 063814 (2014).

\bibitem{article-mul}
J. M. McSloy, W. J. Firth, G. K. Harkness, and G.-L. Oppo, Phys. Rev. E \textbf{66}, 046606 (2002).

\bibitem{bs3}
P. Parra-Rivas, D. Gomila, P. Colet and L. Gelens, Eur. Phys. J. D \textbf{71}, 198 (2017).

\bibitem{bs4}
Y. Wang, F. Leo, J. Fatome, M. erkintalo, S. G. Murdoch, and S. Coen, Optica \textbf{4}, 855 (2017).

\bibitem{ip1}
J. K. Jang, M. Erkintalo, K. Luo, G.-L. Oppo, S. Coen and S. G. Murdoch, New. J. Phys. \textbf{18}, 033034 (2016)

\bibitem{rdf1}
S. Roy and S. K. Bhadra, J. Lightwave. Tech. \textbf{26}, 2301 (2008).

\bibitem{HAELTERMAN1992401}
M. Haelterman, S. Trillo, and S. Wabnitz, Opt. Comm. \textbf{91}, 401 (1992).

\bibitem{PhysRevE.62.8726}
T. Maggipinto, M. Brambilla, G. K. Harkness, and W. J. Firth, Phys. Rev. E \textbf{62}, 8726 (2000).

\bibitem{article00}
V. I. Karpman, and V. V. Solov'ev, Physica D \textbf{3}, 487 (1981).

\bibitem{bs5}
B. A. Malomed, Phys. Rev. A \textbf{44}, 6954 (1991).

\bibitem{PhysRevE.47.2874}
B. A. Malomed, Phys. Rev. E \textbf{47}, 2874 (1993). 

\bibitem{Smith:94}
N. J. Smith, W. J. Firth, K. J. Blow, and K. Smith, Opt. Lett. \textbf{19}, 16 (1994).

\bibitem{bs1}
A. V. Buryak and N. N. Akhmediev, Phys. Rev. E \textbf{51}, 3572 (1995).

\bibitem{Maruta:95}
A. Maruta and Y. Kodama, Opt. Lett. \textbf{20}, 1752 (1995).

\bibitem{article_pp}
A. Hause, H. Hartwig, M. Bohm, and F. Mitschke, Phys. Rev. A \textbf{78}, 63817 (2008).

\bibitem{article_con_bs1}
V. V. Afanasjev, P. L. Chu and B. A. Malomed, Phys. Rev. E \textbf{57}, 1088 (1998).

\bibitem{Parmar:17}

G. S. Parmar, S. Jana and B. A. Malomed, J. Opt. Soc. Am. B \textbf{34}, 850 (2017).

\bibitem{app8020201}
L. Gui, P. Wang, Y. Ding, K. Zhao, C. Bao, X. Xiao, and C. Yang, Appl. Sci. \textbf{8}, 201 (2018).

\bibitem{PhysRevA.97.013816}

A. G. Vladimirov, S. V. Gurevich, and M. Tlidi, Phys. Rev. A \textbf{97}, 013816 (2018).

\bibitem{prl}
D. Turaev, A. G. Vladimirov, and S. Zelik, Phys. Rev. Lett. \textbf{108}, 263906 (2012).

\bibitem{ic2}

P. Wang, X. Xiao, and C. Yang, Opt. Lett. \textbf{42}, 29 (2017).


\end{thebibliography}

\end{document}